\documentclass[useAMS,usenatbib,usegraphicx]{mn2e}
\usepackage{amsmath}
\usepackage{mathtools} 
\usepackage[T1]{fontenc} 
\usepackage{times}
\usepackage{bm}          
\usepackage{flushend} 
\newcommand{\mr}[1]{\mathrm{#1}}
\newcommand{\mat}[1]{\bm{\mathrm{#1}}}

\title[Improving lognormal models for cosmological fields]{Improving lognormal models for cosmological fields}
\author[H. S. Xavier et al.]
{Henrique S. Xavier$^{1,2}$\thanks{E-mail: hsxavier@if.usp.br}, Filipe B. Abdalla$^{2,3}$ and Benjamin Joachimi$^2$\\
$^{1}$ Instituto de Astronomia, Geof\'{i}sica e Ci\^{e}ncias Atmosf\'{e}ricas, Universidade de S\~{a}o Paulo, Rua do Mat\~{a}o, 1226, S\~{a}o Paulo, SP 05508-090, Brazil\\
$^{2}$ Department of Physics \& Astronomy, University College London, 3rd Floor, 132 Hampstead Road, London  NW1 2PS, UK\\
$^{3}$ Department of Physics and Electronics, Rhodes University, PO Box 94, Grahamstown, 6140, South Africa}

\begin{document}

\maketitle

\begin{abstract}

It is common practice in cosmology to model large-scale structure observables as lognormal random fields, 
and this approach has been successfully applied in the past to the matter density and weak lensing 
convergence fields separately. We argue that this approach has fundamental limitations which prevent 
its use for jointly modelling these two fields since the lognormal 
distribution's shape can prevent certain correlations to be attainable. Given the need of ongoing and future 
large-scale structure surveys for fast joint simulations of clustering and weak lensing, we propose 
two ways of overcoming these limitations. The first approach slightly distorts the power 
spectra of the fields using one of two algorithms that minimises either the absolute or the 
fractional distortions. The second one is by obtaining more accurate convergence marginal distributions, 
for which we provide a fitting function, by integrating the lognormal density along the line of sight. 
The latter approach also provides a way to 
determine directly from theory the skewness of the convergence distribution and, therefore, the parameters 
for a lognormal fit. We present the public code \emph{Full-sky Lognormal 
Astro-fields Simulation Kit} ({\sc flask}) which can make tomographic realisations on the sphere 
of an arbitrary number of correlated lognormal or Gaussian random fields by applying either of the 
two proposed solutions, and show that it can create joint simulations of clustering and lensing with 
sub-per-cent accuracy over relevant angular scales and redshift ranges.   
\end{abstract}


\section{Introduction}
\label{sec:Introduction}

One important concept used in cosmology is the random field, i.e. a field defined in space $V$
whose value $F(\bm{r})$ at position $\bm{r}$ is a random variable \citep[see][]{Peebles93x}. Examples of cosmological 
random fields are the matter density, matter velocity, CMB temperature fluctuations and polarisation, 
gravitational lensing convergence and shear fields. The full characterisation of a random field could be 
obtained with the specification of the joint probability density function (PDF) $f_{\mr{joint}}(\bm{F})$ for 
$\bm{F}=\{F(\bm{r})\, |\, \bm{r} \in V\}$.

A common and simple approximation used is to assume that $f_{\mr{joint}}(\bm{F})$ is a multivariate Gaussian 
distribution. In this scenario, all marginal distributions -- the PDFs $f[F(\bm{r})]$ for any particular 
$\bm{r}$ -- are Gaussians and $f_{\mr{joint}}$ is fully characterised by the mean vector 
$\bm{\mu}=\{ \mu(\bm{r})\, |\, \bm{r} \in V\}$ (which in cosmology is generally zero) 
and the covariance matrix $\mat{C}(\bm{r}, \bm{r'})$. Within this model it is possible to 
fully characterise $f_{\mr{joint}}(\bm{F})$ (and therefore the random field) by constraining 
$\mat{C}(\bm{r}, \bm{r'})$, and probably the simplest way to do this is to measure 
the field's correlation function $\xi_{\mr{F}}(\bm{r}, \bm{r'})$ 
[which for a zero mean Gaussian field is actually equal to $\mat{C}(\bm{r}, \bm{r'})$] or its counterpart in Fourier or harmonic 
space. Further simplifications come from the statistical homogeneity of the Universe, 
which makes $\xi_{\mr{F}}(\bm{r}, \bm{r'})=\xi_{\mr{F}}(\bm{r}-\bm{r'})$, and statistical isotropy, 
which leads to $\xi_{\mr{F}}(\bm{r}-\bm{r'})=\xi_{\mr{F}}(|\bm{r}-\bm{r'}|)$.

In some cases the multivariate Gaussian distribution is clearly not a good approximation. The matter density 
contrast $\delta(\bm{r})=[\rho(\bm{r})-\bar{\rho}]/\bar{\rho}$, where $\rho(\bm{r})$ is the density at position 
$\bm{r}$ and $\bar{\rho}$ its average, and the lensing convergence $\kappa(\bm{r})$ 
marginal distributions have hard lower limits which are not obeyed by Gaussian distributions and they 
show significant skewnesses and heavy tails at large values. A better approximation for 
$f_{\mr{joint}}(\bm{F})$ is the multivariate shifted lognormal distribution 
\citep{Coles91mn, Taruya02mn, Hilbert11mn}.

If a set of variables follows a multivariate lognormal distribution, this means that their logarithms 
follow a multivariate Gaussian distribution. The ``shifted'' term expresses simply that the distribution 
is translated around the space populated by the variables (see Sec. \ref{sec:lognormal-variables} for 
details). Even though this model introduces the shifts 
as extra parameters, they are in principle fixed by theory so a measurement of $\xi_{\mr{F}}(\bm{r}, \bm{r'})$ 
would also fully determine $f_{\mr{joint}}(\bm{F})$; if the shifts are left to vary, an extra measurement 
like the marginals' skewnesses are needed. This model has been extensively used for representing both the 
matter/galaxy densities \citep{Coles91mn, Chiang13mn} and the convergence field \citep{Taruya02mn, Hilbert11mn}, 
and it was shown to provide a better approximation than the multivariate Gaussian, but it was also shown 
to depart from observational results and numerical simulations 
\citep{Kofman94mn, Bernardeau95mn, Kayo01mn, Joachimi11mn, Neyrinck11mn, Seo12bmn}. 
One of its main uses is to quickly simulate large-scale structure (LSS) observations to estimate measurement 
errors \citep{Chiang13mn, Alonso15mn} and to test pipelines and estimators \citep[e.g.][]{Beutler14mn}, 
all crucial steps for LSS surveys like 
the Dark Energy Survey\footnote{\texttt{\footnotesize{http://www.darkenergysurvey.org}}} \citep[DES;][]{DES05mn}, 
\emph{Euclid}\footnote{\texttt{\footnotesize{http://www.euclid-ec.org}}} \citep{Lumb09x}, 
the Javalambre Physics of the accelerating universe Astrophysical 
Survey\footnote{\texttt{\footnotesize{http://j-pas.org}}} \citep[J-PAS;][]{Benitez14mn}, 
the Large Synoptic Survey Telescope\footnote{\texttt{\footnotesize{http://www.lsst.org}}} \citep[LSST;][]{LSST09mn} and 
the \emph{Wide-field Infrared Survey Explorer}\footnote{\texttt{\footnotesize{http://wise.ssl.berkeley.edu}}} 
\citep[\emph{WISE};][]{Wright10mn}. 
Note that all these projects will cover large portions of the sky (from 5000 $\mr{deg^2}$ onwards) 
and many will reach redshifts of one or more.  

In this paper we highlight an intrinsic limitation of lognormal variables mostly unknown 
by the astrophysical community that might irremediably prevent its use for modelling 
simultaneous measurements of galaxy density and weak lensing. This limitation comes from 
three combined facts: (a) the relation between two lognormal variables with different 
skewnesses is non-linear; (b) in cosmology the widely used measure for dependence between 
two variables is the Pearson correlation coefficient which works well only for linearly related 
variables; (c) the assumption that the density is lognormally distributed means 
that the convergence is not. We propose two different approaches to deal with this issue: 
distorting either the density and convergence fields' auto- and cross-power spectra or the 
convergence marginal distributions (away from lognormals, making it in fact more realistic). 
We also present the open source code entitled Full-sky Lognormal Astro-fields Simulation Kit 
({\sc flask})\footnote{\texttt{\footnotesize{http://www.astro.iag.usp.br/\char`~flask}}}, 
capable of creating tomographic Gaussian and lognormal realisations of multiple 
correlated fields (multiple tracers, weak lensing convergence, etc.) on the full sky -- using 
spherical coordinates -- and of applying the two corrections suggested above.

This paper is organised as follows: an introduction to lognormal variables is given 
in Sec. \ref{sec:lognormal-variables}, then in Sec. \ref{sec:toy-model} we show how 
items (a) and (b) referred to above combine in a way to restrict the covariance matrices 
realisable by lognormal variables, while in Sec. \ref{sec:harmonic-space} we analyse how 
this restriction translates into the harmonic space (i.e. the computation of angular power 
spectra). We then show in Sec. \ref{sec:consequences} 
that the density lognormality assumption leads to a non-lognormal distribution for the weak 
lensing convergence field, which might cause the lognormal model failure. Nevertheless, using 
this assumption we derive an analytical way of computing the convergence lognormal shift 
parameter which can be used to model the convergence field alone without resorting to 
ray-tracing measurements in $N$-body simulations; in this section we also present a fitting 
function for the convergence distribution to better describe its deviations from the lognormal model. 
Sec. \ref{sec:solutions} details the 
two solutions to the modelling problem, already hinted by the previous sections: distorting 
the power spectra or using a theoretically consistent distribution for the convergence. 
Our code {\sc flask} is described in 
Sec. \ref{sec:code-description}, with an overview given in Sec. \ref{sec:code-overview} and the 
details in Sec. \ref{sec:code-details}. We conclude and summarise our work in Sec. \ref{sec:conclusions}.

\section{Lognormal variables}

\subsection{Definition and properties}
\label{sec:lognormal-variables}

\begin{figure}
  \includegraphics[width=1\columnwidth]{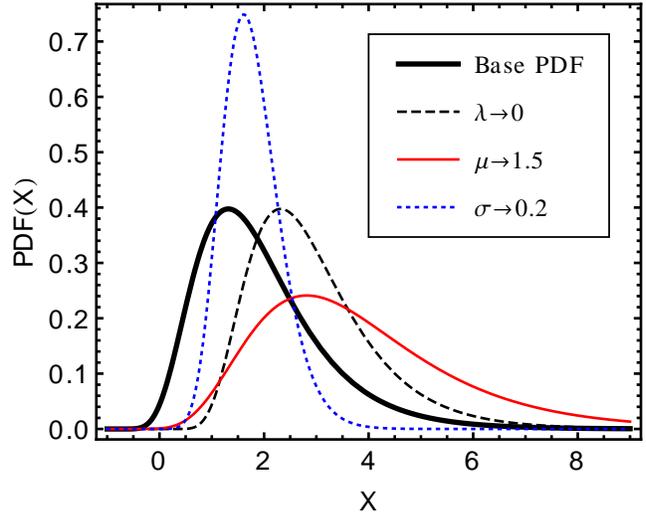}
  \caption{Examples of lognormal probability density functions (PDFs). The base distribution 
(presented in thick black line) has the parameters $\mu=1$, $\sigma=0.4$ and $\lambda=1$. The 
remaining curves are obtained by changing the value of one parameter at a time. The $\lambda$ change 
to zero translates the curve to the right (dashed black line), the $\mu$ change to 1.5 stretches the 
curve to the right (solid red line) and the decrease in $\sigma$ to 0.2 changes the distribution's 
shape, making it less skewed and closer to a Gaussian (dotted blue line).}
  \label{fig:ln-pdfs} 
\end{figure}

Given a set of variables $Z_i$ following a multivariate Gaussian distribution with 
mean vector elements $\mu_i$ and covariance matrix elements $\xi_{\mr{g}}^{ij}$, we call 
the random variables:

\begin{equation}
X_i=e^{Z_i}-\lambda_i
\label{eq:def-lognormal-variable} 
\end{equation}
multivariate shifted lognormal variables, or lognormal variables for short in this paper. 
The parameters $\lambda_i$ are called ``shifts'' by \citet{Hilbert11mn} while $-\lambda_i$ 
are called ``minimum values'' by \citet{Taruya02mn} and ``thresholds'' in the statistics 
literature \citep[e.g.][]{Crow88x}. A single lognormal variable is then fully described by 
three parameters: the shift $\lambda_i$ (which acts as a location parameter), the associated 
Gaussian variable's mean $\mu_i$ (which acts as a scaling parameter) and the associated 
Gaussian variable's variance $\sigma_i^2\equiv \xi_g^{ii}$ (which acts as a shape parameter). Since 
it possesses one extra parameter in comparison to Gaussian variables, it is more flexible than 
the latter. In fact, it tends to the Gaussian case as $\sigma_i^2\rightarrow 0$. Fig. \ref{fig:ln-pdfs} 
presents examples of the effects on the PDF of changing the distribution's parameters.

The relations between the correlation function of lognormal variables and their parameters 
$\mu_i$, $\lambda_i$ and $\xi_{\mr{g}}^{ij}$ have been presented in the astrophysical literature for 
the case $\lambda_i=0$ by \citet{Coles91mn} and for $\langle X_i\rangle =0$ and $\lambda_i=\lambda$ 
by \citet{Hilbert11mn}. Here we generalise their results for multiple arbitrary shifts $\lambda_i$ and 
expected values (i.e. statistical ensemble averages) $\langle X_i\rangle$.

The mean value $\langle X_i\rangle$ of a lognormal variable $X_i$ can be obtained from Eq. 
\ref{eq:def-lognormal-variable} by expanding the exponential as an infinite series:
\begin{equation}
\langle e^{Z_i} \rangle = e^{\mu_i}\langle e^{Z_i-\mu_i} \rangle=  e^{\mu_i}\sum_{n=0}^\infty\frac{\langle (Z_i-\mu_i)^n\rangle}{n!}
\label{eq:e-series}
\end{equation}
and remembering that the Gaussian central moments $\langle (Z_i-\mu_i)^n\rangle$ follow the relation 
\begin{equation}
\langle (Z_i-\mu_i)^n\rangle = \begin{cases}
0, & \text{if $n$ is odd,}\\
\frac{n!}{(n/2)!}\left(\frac{\sigma_i^2}{2}\right)^{n/2}, & \text{if $n$ is even.}\\
\end{cases}
\label{eq:gaussian-moments}
\end{equation}
By inserting Eq. \ref{eq:gaussian-moments} into Eq. \ref{eq:e-series} and defining a new 
summation index $m\equiv n/2$, we get:

\begin{equation}
\langle X_i \rangle = e^{\mu_i+\frac{\sigma_i^2}{2}} - \lambda_i.
\label{eq:lognormal-mean}
\end{equation}

To derive the relation between the lognormal and associated Gaussian covariances 
$\xi_{\mr{ln}}^{ij}$ and $\xi_{\mr{g}}^{ij}$, we can write $Z_i$ as a sum of zero-mean independent 
Gaussian variables $g_n$, e.g. $Z_1=\mu_1+g_1+g_0$ and $Z_2=\mu_2+g_2+g_0$ such that 
$\langle (Z_1-\mu_1)(Z_2-\mu_2) \rangle = \langle g_0^2 \rangle$. This allows us to treat the expectation value 
$\langle e^{Z_1-\mu_1}e^{Z_2-\mu_2} \rangle$ as a product of independent terms:
\begin{equation}
\langle e^{Z_1-\mu_1}e^{Z_2-\mu_2} \rangle = \langle e^{g_1} \rangle\langle e^{g_2} \rangle\langle e^{2g_0} \rangle
\label{eq:exp-product-mean}
\end{equation}
to which we can apply the same procedure used to derive Eq. \ref{eq:lognormal-mean}, leading to:

\begin{equation}
\xi_{\mr{ln}}^{ij}\equiv\langle X_i X_j \rangle -\langle X_i\rangle\langle X_j\rangle= \alpha_i\alpha_j(e^{\xi_{\mr{g}}^{ij}} - 1),
\label{eq:lognormal-cov}
\end{equation}

\begin{equation}
\xi_{\mr{g}}^{ij}= \ln\left(\frac{\xi_{\mr{ln}}^{ij}}{\alpha_i\alpha_j}+1\right),
\label{eq:gaussian-cov}
\end{equation}
where $\alpha_i\equiv \langle X_i \rangle + \lambda_i>0$. Again the same method can be used to derive a relation 
for the three-point correlation function of lognormal variables:

\begin{equation}
\begin{split}
\zeta_{\mr{ln}}^{ijk}&\equiv \langle (X_i-\langle X_i\rangle)(X_j-\langle X_j\rangle)(X_k-\langle X_k\rangle)\rangle = \\
&=\frac{\xi_{\mr{ln}}^{ij}\xi_{\mr{ln}}^{jk}\xi_{\mr{ln}}^{ki}}{\alpha_i\alpha_j\alpha_k}
+\frac{\xi_{\mr{ln}}^{ji}\xi_{\mr{ln}}^{ik}}{\alpha_i}+\frac{\xi_{\mr{ln}}^{ij}\xi_{\mr{ln}}^{jk}}{\alpha_j}
+\frac{\xi_{\mr{ln}}^{ik}\xi_{\mr{ln}}^{kj}}{\alpha_k}.
\end{split}
\label{eq:lognormal-3pt}
\end{equation}

By setting all indices in Eqs. \ref{eq:lognormal-cov} and \ref{eq:lognormal-3pt} to the same value, we get relations 
for the variance $v_i$ and skewness $\gamma_i$ of a lognormal variable:

\begin{equation}
v_i\equiv\langle X_i^2 \rangle - \langle X_i \rangle^2 = \alpha_i^2(e^{\sigma_i^2}-1),
\label{eq:lognormal-var}
\end{equation}

\begin{equation}
\gamma_i\equiv\frac{\langle (X_i -\langle X_i\rangle)^3\rangle}{v_i^{3/2}} = 
\frac{\sqrt{v_i}}{\alpha_i}\left( \frac{v_i}{\alpha_i^2}+3 \right).
\label{eq:lognormal-skew}
\end{equation}

The equation above can be inverted to obtain $\alpha_i$ as a function of $\gamma_i$ 
and $v_i$; although in principle Eq. \ref{eq:lognormal-skew} admits more than 
one $\alpha_i$ as a solution, only one of them is real as the relation is monotonic. 
The shift parameter $\lambda_i$ can then be written in terms of the variable's mean, 
variance and skewness:

\begin{equation}
\lambda_i = \frac{\sqrt{v_i}}{\gamma_i}\left[ 1+y(\gamma_i)+\frac{1}{y(\gamma_i)}  \right] - \langle X_i \rangle, 
\label{eq:moments2shift}
\end{equation}
\begin{equation}
y(\gamma) \equiv \sqrt[\leftroot{-1}\uproot{2}\scriptstyle 3]{\frac{2+\gamma^2+\gamma\sqrt{4+\gamma^2}}{2}}.
\label{eq:skewness2y}
\end{equation}
Once we have computed $\lambda_i$, we can get the remaining parameters of the lognormal 
distribution that possess the specified first three moments by inverting Eqs. 
\ref{eq:lognormal-mean} and \ref{eq:lognormal-var}:

\begin{equation}
\mu_i = \ln\left( \frac{\alpha_i^2}{\sqrt{\alpha_i^2+v_i}} \right),
\label{eq:mu-from-moments}
\end{equation}

\begin{equation}
\sigma_i = \sqrt{\ln\left( 1+\frac{v_i}{\alpha_i^2} \right)}.
\label{eq:sigma-from-moments}
\end{equation}
This provides us with a method to fit a lognormal distribution to a dataset that exactly 
reproduces its mean, variance and skewness.

\subsection{Intrinsic limitations of multivariate lognormals}
\label{sec:toy-model}

\begin{figure}
  \includegraphics[width=1\columnwidth]{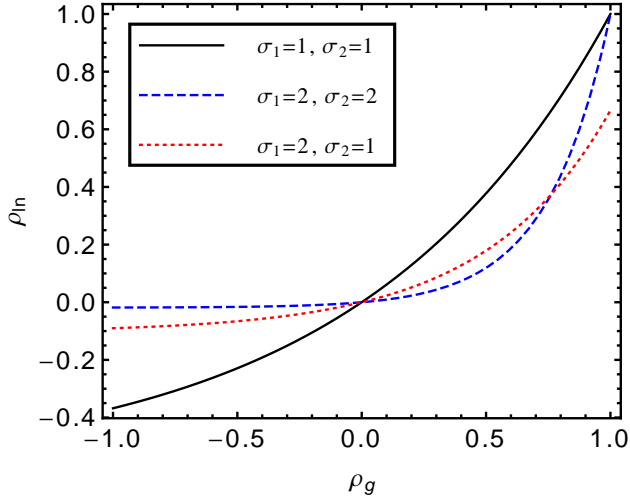}
  \caption{Relationship between the correlation $\rho_{\mr{g}}$ of two Gaussian variables and the 
    correlation $\rho_{\mr{ln}}$ of their associated lognormal variables. The amount of Pearson 
    correlation and anti-correlation of lognormal variables is smaller than the correlation of their 
    Gaussian counterparts and the relation depends on the Gaussian variances $\sigma_1^2$ and $\sigma_2^2$.}
  \label{fig:g-ln-correlation} 
\end{figure}

\begin{figure}
  \includegraphics[width=1\columnwidth]{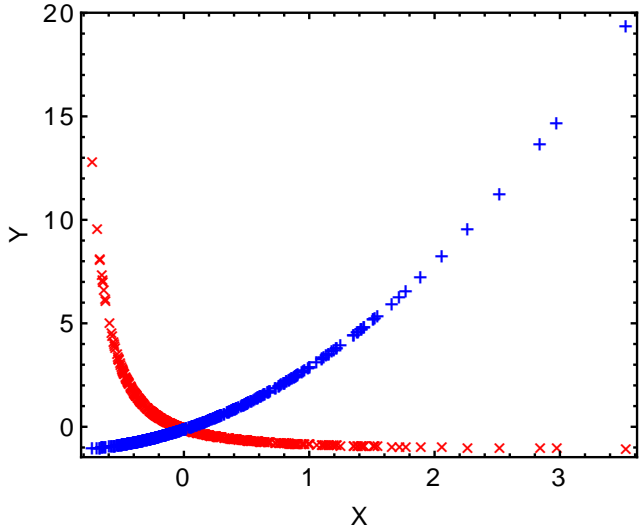}
  \caption{Each marker shows 400 realisations of fully dependent lognormal random variables $X$ and $Y$, 
    that is, $\ln(Y+1)=-2\ln(X+1)$, shown as red crosses, and $\ln(Y+1)=2\ln(X+1)$, shown as 
    blue plus signs. Even though their associated Gaussian variables are completely anti-correlated 
    and completely correlated, respectively, the absolute values of their Pearson correlations are 
    smaller than one: 0.58 and 0.94. This happens because the relation between them is non-linear.}
  \label{fig:x-y-lnvar} 
\end{figure}

To expose the fundamental limitations that lognormal variables face when modelling correlated data, 
consider a toy model consisting of only two variables. We can use Eq. \ref{eq:lognormal-cov} to 
build a relation between the Pearson correlation coefficients of the lognormal variables, $\rho_{\mr{ln}}$, 
and that of their associated Gaussian variables, $\rho_{\mr{g}}$:

\begin{equation}
  \rho_{\mr{ln}}= \frac{e^{\rho_{\mr{g}}\sigma_1\sigma_2}-1}{\sqrt{(e^{\sigma_1^2}-1)(e^{\sigma_2^2}-1)}},
  \label{eq:g-ln-correlation}
\end{equation}
where $\sigma_1^2$ and $\sigma_2^2$ are the variances of the Gaussian variables and serve as shape 
parameters (which fully determines the skewness) of the lognormal distributions. 
The relation above is presented in Fig. \ref{fig:g-ln-correlation} for different values of 
$\sigma_1^2$ and $\sigma_2^2$, where it is possible to note that even perfectly correlated 
Gaussian variables ($\rho_{\mr{g}}=1$) may not result in perfectly correlated lognormal variables. This happens 
because one cannot impose a linear relation between two variables $X$ and $Y$ 
if their distributions have different shapes (e.g. different skewnesses) since such a relation only corresponds to shifting and 
rescaling one distribution to match the other (see Fig. \ref{fig:x-y-lnvar}). These limits on the Pearson correlation 
coefficient can be written in terms of the parameters of lognormal variables:

\begin{equation}
  \frac{\alpha_1\alpha_2}{\sqrt{v_1v_2}}\left( e^{-L}-1\right) <\rho_{\mr{ln}}<
  \frac{\alpha_1\alpha_2}{\sqrt{v_1v_2}}\left( e^{L}-1\right)\text{, with}
  \label{eq:ln-corr-limits}
\end{equation}

\begin{equation}
  L \equiv \sqrt{\ln\left( \frac{v_1}{\alpha_1^2} +1\right)\ln\left( \frac{v_2}{\alpha_2^2} +1\right)}.
  \label{eq:L-def}
\end{equation}

A more rigorous and general (but also complex) proof of the correlation limits above can be 
obtained from the use of copulas \citep{Nelsen06x}: any multivariate distribution can be described by 
a copula -- a multidimensional function that alone specifies the variables' inter-dependences -- together with the 
one-dimensional marginal distributions of these variables. Copulas are useful because the dependence between the 
random variables becomes detached from their marginal distributions. The Fr\'{e}chet--Hoeffding theorem 
states that all copulas are limited by specific functions $W$ and $M$ called \emph{lower} and 
\emph{upper Fr\'{e}chet--Hoeffding bounds}. It is then possible to derive Eq. \ref{eq:ln-corr-limits} 
by setting the two-dimensional copula to $W$ and $M$ and calculating the resulting correlations
\citep[see also][]{Denuit03mn}.

Suppose now that one ignores the limits above and assigns to a pair of lognormal variables a 
valid (i.e. positive-definite) covariance matrix but that violates Eq. \ref{eq:ln-corr-limits}. 
By using Eq. \ref{eq:gaussian-cov} one would find 
$|\rho_{\mr{g}}|>1$, which for a $2\times 2$ covariance matrix corresponds to being invalid 
(i.e. non-positive-definite).\footnote{Assuming that the diagonal terms are positive.} 
Since lognormal variables are associated to Gaussian variables by definition, the 
non-positive-definiteness of the Gaussian variables' covariance matrix shows that such 
lognormal variables cannot exist. In other words, a covariance matrix for lognormal variables 
is only valid if both itself and its Gaussian counterpart are positive-definite.\footnote{Since 
the relation between Gaussian variables is always linear, their Pearson correlation actually 
reflects the degree of dependence between them.} 
This statement can be extended to covariance matrices of arbitrary size; this is important because, 
when dealing with more than two variables, the condition set 
by Eq. \ref{eq:ln-corr-limits} is necessary but not sufficient. 

As an example of what may happen in more complex cases, imagine there are three lognormal variables 
with $\alpha_1=1$ and $\alpha_2=\alpha_3=0.1$ that follow the covariance 
matrix below on the left:  
\begin{equation}
\left(\begin{array}{ccc}
1    & 0.45 & 0.45 \\
0.45 & 1    & 0.40 \\
0.45 & 0.40 & 1    \\
\end{array}\right)
\rightarrow
\left(\begin{array}{ccc}
1    & 0.95 & 0.95 \\
0.95 & 1    & 0.80 \\
0.95 & 0.80 & 1    \\
\end{array}\right).
\label{eq:3x3-matrix-ex}
\end{equation}
This positive-definite (and seemingly innocent) covariance matrix hides the fact that the 
dependence between the first variable and the two others is very strong (the maximum correlation 
allowed by the difference in the shape of their distributions is $\sim0.50$) whereas the dependence between the 
last two variables is not strong enough to be compatible with the former (since they have the same shape, 
their maximum correlation is $1$). Indeed, the correlation matrix of the associated Gaussian variables (right-hand side of 
Eq. \ref{eq:3x3-matrix-ex}) is non-positive-definite.

\begin{figure*}
  \includegraphics[width=1\textwidth]{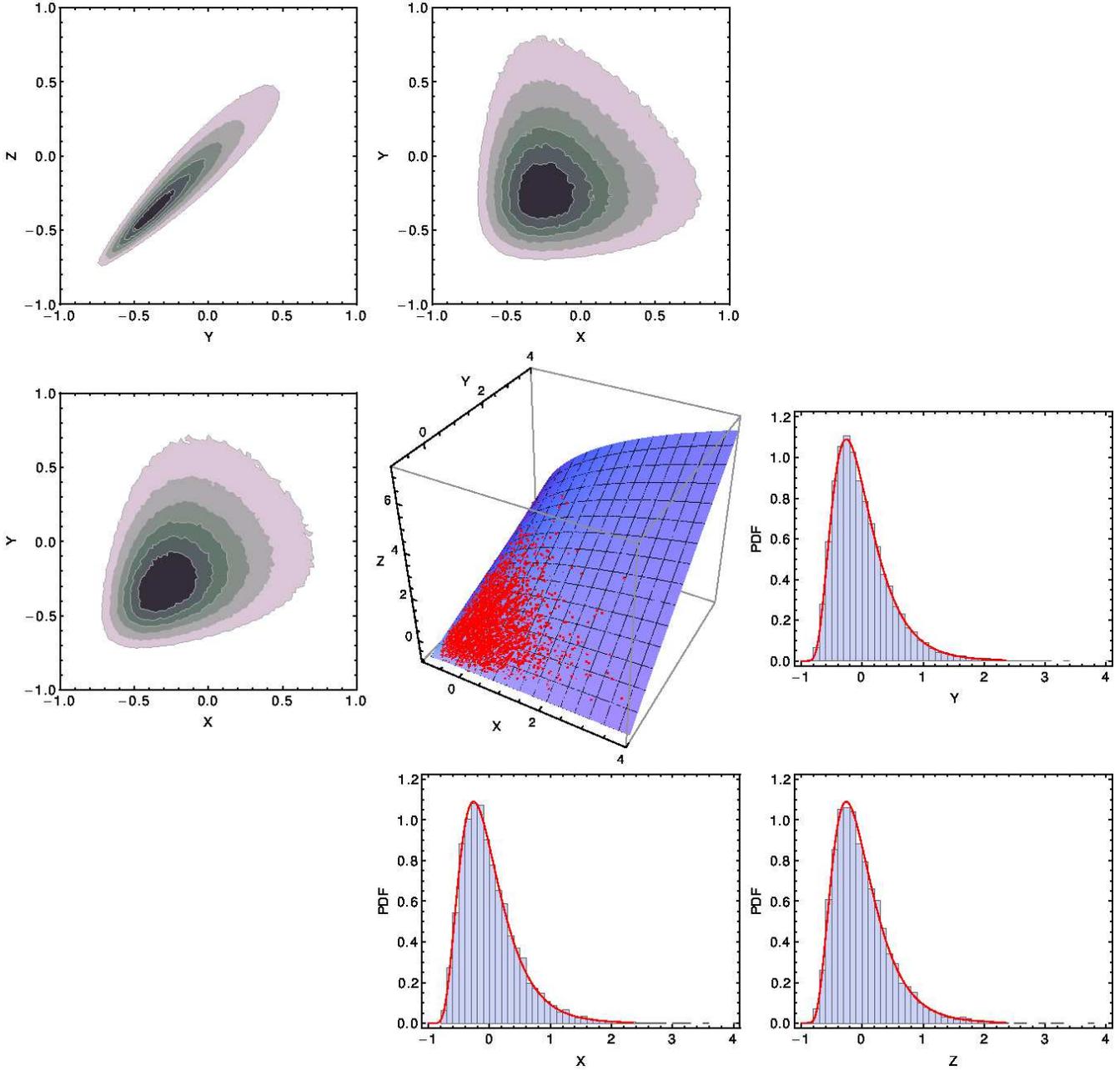}
  \caption{The plots show the three 1D marginal distributions (bottom right corner), three 
    2D marginal distributions (top left corner) and the scatter plot (centre) of three 
    correlated lognormal random variables (yellow dots). The variables have a strong non-linear 
    dependence that confines them into a space of lower dimensionality (the curved surface 
    shown in the centre plot) and therefore stronger Pearson correlations might not be 
    achievable, even though the 2D marginals do not indicate that and the 1D 
    marginals have exactly the same shape. The correlations between the variables (together with their variances 
    and minimum values) determine the 3D distribution's shape.}
  \label{fig:3D-plots} 
\end{figure*}

The limitations over three or more lognormal variables appear even when the one-dimensional marginals 
are exactly the same. Fig. \ref{fig:3D-plots} shows that the Gaussian covariances serve 
as shape parameters for the multivariate lognormal distribution just as the Gaussian variances 
$\sigma_i$ do due to the non-linearity of the transformation. The shape of the distribution can be 
such that projections on two-dimensional spaces might give the impression that tighter correlations 
are possible when they are not.

Another way to deduce the connection between lognormal variables and their Gaussian 
counterparts covariance matrix is the following:
\begin{enumerate}
\item \label{it:ln-g-pair}
  \emph{Fact:} lognormal variables always have, by definition, Gaussian variables associated to them;
\item \label{it:cov-m-needed}
  \emph{Fact:} any set of $N$ random variables must have a valid $N\times N$ covariance matrix associated to it;
\item \label{it:hypothesis}
  \emph{Hypothesis:} $\{X_1, ..., X_N\}$ is a set of multivariate lognormal variables and it has the 
  covariance matrix $\mat{C_{\mr{ln}}}$;
\item \emph{Consequence from \ref{it:ln-g-pair}:} there is a set $\{Z_1, ..., Z_N\}$ 
  of Gaussian variables related to $\{X_1, ..., X_N\}$ by Eq. \ref{eq:def-lognormal-variable};
\item \label{it:conclusion}
  \emph{Consequence from \ref{it:cov-m-needed}:} $\{Z_1, ..., Z_N\}$ have a valid covariance matrix $\mat{C_{\mr{g}}}$ 
  that can be obtained from Eq. \ref{eq:lognormal-cov}.
\end{enumerate}
If our final conclusion \ref{it:conclusion} is not true, our hypothesis \ref{it:hypothesis} 
must be false, that is, either $\{X_1, ..., X_N\}$ are not multivariate lognormal variables or they do 
not follow $\mat{C_{\mr{ln}}}$. Trying to enforce both at the same time would be like requesting two different 
angles from an equilateral triangle. Note that the relation between $\mat{C_{\mr{ln}}}$ and $\mat{C_{\mr{g}}}$ depends on the 
full multi-dimensional PDF of $\{X_1, ..., X_N\}$ so although it might not be a multivariate lognormal 
it can retain, in principle, marginal lognormal distributions. 

\subsection{Limitations in harmonic space}
\label{sec:harmonic-space}

A collection of 3D isotropic random fields can be described by a set of angular correlation functions 
$\xi^{ij}(\theta)$ for fields and redshift slices specified by the indices $i$ and $j$. These correlation 
functions can be expressed in terms of angular power spectra $C^{ij}(\ell)$ through the relations 
\citep{Durrer08x}:

\begin{equation}
  C^{ij}(\ell) = 2\pi \int_0^\pi \xi^{ij}(\theta)P_{\ell}(\cos\theta) \sin\theta\mr{d}\theta,
  \label{eq:legendre-trafo-cl}
\end{equation}

\begin{equation}
  \xi^{ij}(\theta) = \frac{1}{4\pi}\sum_{\ell=0}^\infty(2\ell+1)C^{ij}(\ell)P_{\ell}(\cos\theta),
  \label{eq:legendre-trafo-xi}
\end{equation}
where $P_{\ell}(\mu)$ are Legendre polynomials.

If the fields in question follow lognormal distributions in real space, the relation between 
the angular power spectra $C_{\mr{ln}}^{ij}(\ell)$ and $C_{\mr{g}}^{ij}(\ell)$ that describe the 
lognormal fields and their associated Gaussian counterparts, respectively, is

\begin{equation}
  C_{\mr{g}}^{ij}(\ell) = 2\pi\int_{-1}^{1}\ln\left[ 
    \sum_{\ell'=0}^\infty \frac{(2\ell'+1)}{4\pi}\frac{C_{\mr{ln}}^{ij}(\ell')}{\alpha_i\alpha_j}P_{\ell'}(\mu)+1\right] 
  P_\ell(\mu)\mr{d}\mu.
  \label{eq:cl-g-ln}
\end{equation}
Although the relation above is not as direct as the one in real space (see Eq. \ref{eq:gaussian-cov}), it 
takes advantage of isotropy to make each multipole independent of one another and reduce the 
dimensionality of the covariance matrices to the number of fields and redshift slices specified 
by $i$ and $j$. In other words, for each $\ell$ we have an independent covariance matrix 
$\mat{C}(\ell)$ with elements $C^{ij}(\ell)$.

\begin{figure}
  \includegraphics[width=1\columnwidth]{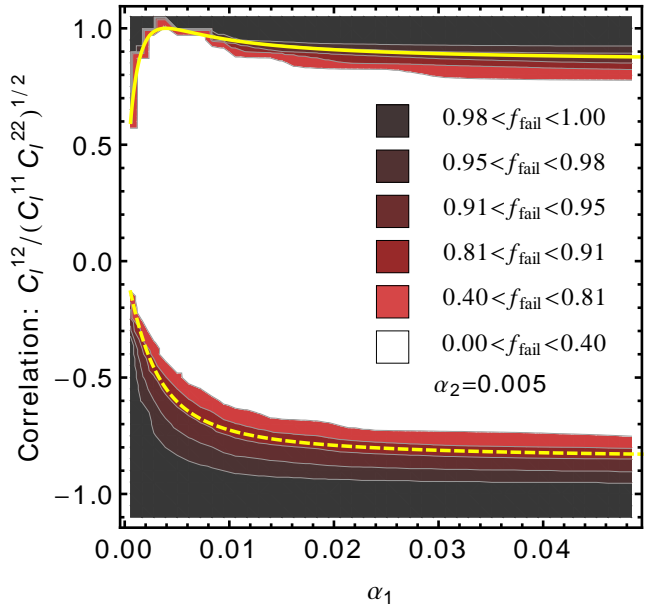}
  \caption{The dashed and solid thick yellow lines show the lower and upper correlation limits according to 
    Eq. \ref{eq:ln-corr-limits}; we set $\alpha_2=0.005$ and the variances $v_1$ and $v_2$ were computed using Eq. 
\ref{eq:legendre-trafo-xi} with $i=j$ and $\theta=0$.  
The shaded regions are coloured according to the fraction $f_{\mr{fail}}$ of $C_{\mr{ln}}^{ij}(\ell)$ in the range 
$2\leq \ell \leq 5000$ that failed to result in positive-definite $C_{\mr{g}}^{ij}(\ell)$ (see the text). In harmonic 
space, the correlation limits get blurred but retain approximately the same form.}
  \label{fig:harmonic-limits} 
\end{figure}

It is difficult to derive analytically how the restriction described in Sec. \ref{sec:toy-model} affects the 
relation between $C_{\mr{ln}}^{ij}(\ell)$ and $C_{\mr{g}}^{ij}(\ell)$: given it is local in real 
space, the relation becomes non-local in harmonic space, i.e. $C_{\mr{g}}^{ij}(\ell)$ relates to a combination 
of $C_{\mr{ln}}^{ij}(\ell')$ with different $\ell'$; moreover, the multipoles described by $C_{\mr{ln}}^{ij}(\ell)$ 
are not themselves lognormal. However, a highly correlated field in real space should 
be highly correlated in harmonic space as well and therefore the conditions set in Sec. \ref{sec:toy-model} 
cannot be completely avoided. This is shown in Fig. \ref{fig:harmonic-limits}.

To draw the shaded regions in Fig. \ref{fig:harmonic-limits} we first computed the convergence auto- and cross-power 
spectra for sources inside 0.1-wide top-hat redshift bins centred at $z_1=0.5$ and $z_2=0.6$ using 
{\sc class}\footnote{\texttt{\footnotesize{http://class-code.net}}} \citep{Blas11mn, Dio13mn} and a flat $\mr{\Lambda CDM}$ model. 
These $C_{\mr{ln}}^{ij}(\ell)$ were transformed to $C_{\mr{g}}^{ij}(\ell)$ using Eq. \ref{eq:cl-g-ln} -- implemented 
by {\sc flask} -- and $\alpha_i$ obtained by \citet{Hilbert11mn} using ray tracing through $N$-body simulations. 
For each $\ell$ we built a $2\times 2$ covariance matrix $\mat{C_{\mr{g}}}(\ell)$ 
which was tested for positive-definiteness. To probe the whole parameter space in Fig. \ref{fig:harmonic-limits} 
we repeated this process several times after rescaling the cross power spectrum and changing $\alpha_1$. 

The final message of Fig. \ref{fig:harmonic-limits} is that the limitations described in Sec. \ref{sec:toy-model} 
manifest themselves in harmonic space and can indeed prevent the realisation of multipoles of lognormal fields, 
showing in these cases that the proposed fields cannot exist. Moreover, this seems to be the only relevant process 
affecting the positive-definiteness of $\mat{C_{\mr{g}}}(\ell)$ -- at least in the simple example shown and aside from much 
smaller numerical errors.  

\section{Lognormal large-scale structure models}
\label{sec:consequences}

\subsection{Quantifying the lognormal failure and distorting $C^{ij}(\ell)$}
\label{sec:distortions}

We investigated if the limitations referred to in the previous section manifest themselves 
in the density and convergence fields. We described the projected matter density contrast 
$\delta$ inside redshift bins and the weak lensing convergence $\kappa$ for sources inside 
those bins as multivariate lognormal variables that obey a set of $C_{\mr{ln}}^{ij}(\ell)$ 
with $i=\{\delta(z_1), ..., \delta(z_n), \kappa(z_1), ..., \kappa(z_n)\}$ and inferred the 
model's validity by checking if the matrices $\mat{C_{\mr{g}}}(\ell)$ with elements given by 
Eq. \ref{eq:cl-g-ln} were positive-definite. 

When a matrix $\mat{C_{\mr{g}}}(\ell)$ turned out to be non-positive-definite, we quantified the 
degree of ``non-positive-definiteness'' by computing the fractional change in the 
matrix elements needed to make it positive-definite. For that we used a multi-dimensional 
gradient to minimise the sum of the absolute values of the negative eigenvalues: by computing the change 
in the negative eigenvalues given a small fractional change in each one of the $N\times N$ 
matrix elements we found a preferential direction in this $N\times N$-dimensional space to distort the 
matrix and applied a small change in this direction; we repeated this process until all eigenvalues were positive. 
Another method to regularise a covariance matrix is to perform an eigendecomposition of the 
matrix [$\mat{C_{\mr{g}}}=\mat{Q\Lambda Q}^{-1}$, where $\mat{Q}$ is a matrix formed by the eigenvectors of 
$\mat{C_{\mr{g}}}$ and $\mat{\Lambda}$ is a diagonal matrix formed by $\mat{C_{\mr{g}}}$ eigenvalues] and set 
the negative eigenvalues to zero. However, this method results in minimal absolute rather 
than minimal fractional changes; more specifically, it is guaranteed to minimise the Frobenius norm -- i.e. the matrix 
elements' quadratic sum -- of the difference between the original and regularised matrices \citep{Higham88mn}. 
We confirmed that the fractional change obtained by our method is indeed smaller than the one obtained 
from the latter, and that they both result in fractional changes of similar magnitude for 
$C_{\mr{g}}^{ij}(\ell)$ not too close to zero. Both regularisation methods can be performed by 
{\sc flask}. The regularised $C_{\mr{g}}^{ij}(\ell)$ can be transformed back into 
$C_{\mr{ln}}^{ij}(\ell)$ to give a set of angular power spectra that would not fail to 
represent lognormal fields.

\begin{figure}
  \includegraphics[width=1\columnwidth]{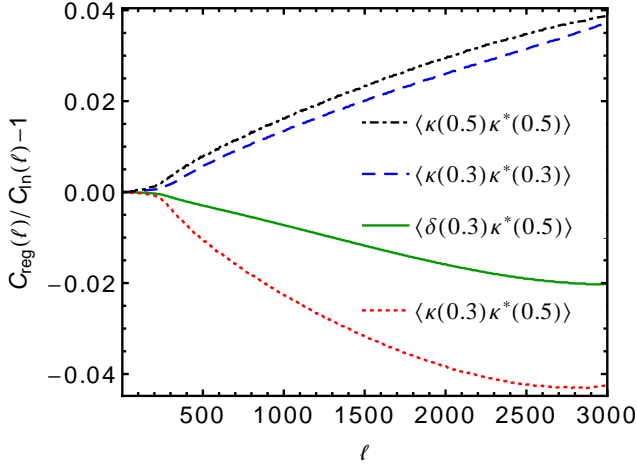}
  \caption{Four highest fractional differences between the original angular power spectra $C_{\mr{ln}}^{ij}(\ell)$ 
    and the regularised one $C_{\mr{reg}}^{ij}(\ell)$ when modelling density and convergence at three redshift bins. 
    The difference increases with $\ell$ up to $\ell=3000$. In general the diagonal terms 
    [auto-$C(\ell)$s] are increased while off-diagonal terms [cross-$C(\ell)$s for different redshift bins] 
    are decreased, reducing the correlation between redshift slices.}
  \label{fig:delta-kappa-3}
\end{figure}

High fractional changes are needed when trying to model both density and convergence 
as lognormal fields. Broadly speaking the amount of $C_{\mr{ln}}^{ij}(\ell)$ distortion required 
to make $C_{\mr{g}}^{ij}(\ell)$ positive-definite increases with $\ell$ and with the 
number of redshift bins, and is higher for the non-linear power spectra computed by {\sc halofit} 
\citep{Smith03mn, Takahashi12mn} and when low-redshift bins are included: with 
the closest bin centred at $z=0.3$, the required amount of change 
goes from $\sim 1.2\%$ ($\sim 4\%$) for 3 bins to $\sim 8\%$ ($\sim 20\%$) for 20 bins 
when using linear (non-linear) power spectra. Other parameters have a smaller impact on 
the fractional changes. As Fig. \ref{fig:delta-kappa-3} shows, changes affect 
mainly the high multipoles and are much larger than the numerical precision expected 
for these operations (see below).

\begin{figure}
  \includegraphics[width=1\columnwidth]{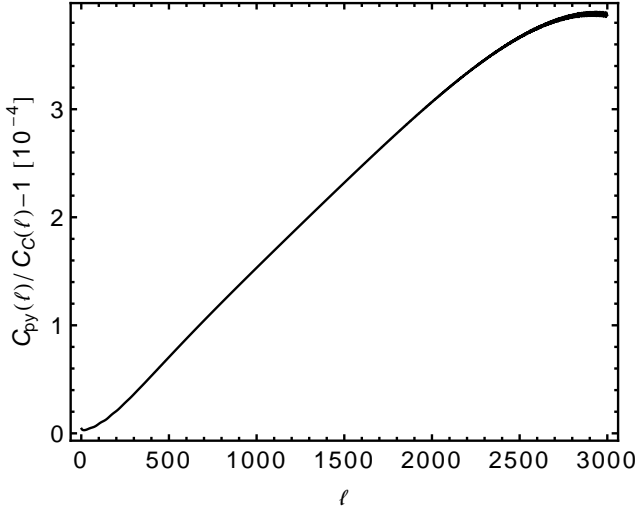}
  \caption{Highest fractional difference between $C_{\mr{g}}^{ij}(\ell)$ computed with 
{\sc python} and {\sc C} routines for density and convergence fields at three different 
redshift bins (total of 21 power spectra). The {\sc python} routine diverges at $\ell\sim\ell_{\mr{max}}$ 
(while the {\sc C} routine does not) so we set $\ell_{\mr{max}}=7000$. Numerical fractional 
errors on the transformation given by Eq. \ref{eq:cl-g-ln} are expected to be smaller 
than $4\times 10^{-4}$ up to $\ell=3000$, specially for the routine in {\sc C}.}
  \label{fig:ratio-c-py}
\end{figure}

To ensure that the results presented in Fig. \ref{fig:delta-kappa-3} 
were not caused by numerical inaccuracies, we used both {\sc class} and 
{\sc camb sources}\footnote{\texttt{\footnotesize{http://camb.info/sources}}} \citep{Challinor11mn} to generate 
the required $C_{\mr{ln}}^{ij}(\ell)$ under a variety of precision settings and performed the 
transformation described in Eq. \ref{eq:cl-g-ln} using two different methods under two different 
programming languages. Our main method (implemented in {\sc flask}) was built in {\sc C} and 
used the discrete Legendre transform coded in {\sc s2kit}\footnote{\texttt{\footnotesize{http://www.cs.dartmouth.edu/\char`~geelong/sphere}}} 
\citep{Kostelec00mn} to go back and forth into harmonic space, while our second method used the 
functions {\sc legval} and {\sc legfit} in {\sc python}'s {\sc numpy} package. Fig. \ref{fig:ratio-c-py} 
shows that numerical fractional errors are expected to remain below $4\times 10^{-4}$. 
We also confirmed the behaviour of our results for Gaussian and top-hat redshift bins of different widths, 
with and without different contributions included in the matter density distribution (redshift 
space distortions, gravitational lensing, Integrated Sachs--Wolfe effect and gravitational 
redshift) and with and without non-linear structure given by {\sc halofit}. 

\begin{figure}
  \includegraphics[width=1\columnwidth]{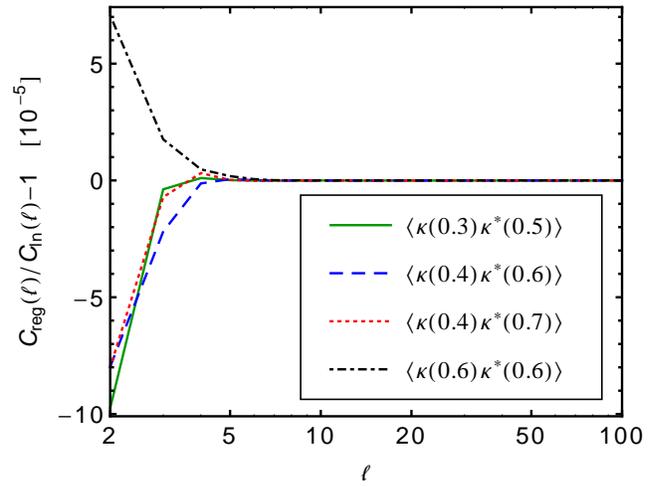}
  \caption{Same as Fig. \ref{fig:kappa-frac-diff} but for convergence only, computed in 19 redshift bins. 
    The difference is largest at low multipoles; we tested up to $\ell=3000$. As before, regularisation  
    reduces the correlations between fields and redshift bins.}
  \label{fig:kappa-frac-diff}
\end{figure}

When considering matter density contrast or weak lensing convergence separately -- i.e. 
when modelling one of these fields independently of the other -- any need for regularising 
covariance matrices results in diminute fractional changes, of order $10^{-5}$, that in general 
affect low multipoles ($\ell \la 50$). Such small deviations from positive-definiteness 
might be caused by numerical inaccuracies and, in any case, are too small to be detectable 
especially at low multipoles where cosmic variance is large. We verified that this pattern 
is maintained for different matter density contributions portfolio, for the linear and non-linear 
power spectra, for {\sc class} and {\sc camb sources} with different precision settings and 
for various redshift ranges and binning (we tested from 2 to 50 redshift bins in the range 
$0.3\la z\la 3.0$). In fact, the {\sc class} computation of density power spectra never resulted 
in non-positive-definite covariance matrices $C_{\mr{g}}^{ij}(\ell)$. As an example, Fig. 
\ref{fig:kappa-frac-diff} shows the four largest fractional differences between the original 
input $C_{\mr{ln}}^{ij}(\ell)$ and the regularised ones $C_{\mr{reg}}^{ij}(\ell)$ when modelling the 
convergence in 19 top-hat redshift bins of width $\Delta z=0.1$ in the range $0.2<z<2.0$.    

\subsection{Density and convergence lognormality inconsistency}
\label{sec:density-ln}

A model where both density and convergence are lognormal variables includes by definition 
an internal inconsistency due to the following connection between the two, which ends 
encoded in the power spectra: one can compute the convergence $\kappa(\bm{\theta},z_{\mr{s}})$ 
for galaxies at angular position $\bm{\theta}$ and redshift $z_{\mr{s}}$ by integrating the 
matter density contrast $\delta(\bm{\theta},z)$ along the line of sight (LoS) 
\citep[eq. 6.16]{Bartelmann01mn}:

\begin{equation}
\kappa(\bm{\theta},z_{\mr{s}}) = \int_0^{z_{\mr{s}}}K(z,z_{\mr{s}})\delta(\bm{\theta},z)\mr{d}z,
\label{eq:bartelmann-kappa}
\end{equation}

\begin{equation}
K(z,z_{\mr{s}}) \equiv \frac{3H_0^2\Omega_{\mr{m}}}{2c^2}
\frac{f[\chi(z)]f[\chi(z_{\mr{s}})-\chi(z)]}{f[\chi(z_{\mr{s}})]}(1+z)\frac{\mr{d}\chi}{\mr{d}z},
\label{eq:wl-kernel}
\end{equation}
where $f(\chi)$ is called \emph{transverse comoving distance}:

\begin{equation}
f(\chi)=\begin{cases}
\frac{1}{\sqrt{-\Omega_{\mr{k}}}}\frac{c}{H_0}\sin\left( \frac{H_0}{c}\sqrt{-\Omega_{\mr{k}}}\chi  \right), & \Omega_{\mr{k}}<0,\\
\chi, & \Omega_{\mr{k}}=0,\\
\frac{1}{\sqrt{\Omega_{\mr{k}}}}\frac{c}{H_0}\sinh\left( \frac{H_0}{c}\sqrt{\Omega_{\mr{k}}}\chi  \right), & \Omega_{\mr{k}}>0
\end{cases}  
\label{eq:transverse-dist}
\end{equation}
and $\chi=\chi(z)$ is the comoving distance, given by 

\begin{equation}
\chi(z) = \frac{c}{H_0}\int_0^z\frac{\mr{d}z'}{E(z')},
\label{eq:radial-dist}
\end{equation}
\begin{equation}
E(z') = \sqrt{\Omega_{\mr{m}}(1+z')^3+\Omega_{\mr{k}}(1+z')^2+\Omega_{\mr{de}}(1+z')^{3(1+w)}}.
\label{eq:hubble-par}
\end{equation}
In these equations $H_0$ is the Hubble's constant, $c$ is the speed of light, $\Omega_{\mr{m}}$, $\Omega_{\mr{de}}$ 
and $\Omega_{\mr{k}}=1-\Omega_{\mr{m}}-\Omega_{\mr{de}}$ are the total matter, dark energy and curvature density 
parameters, respectively, and $w$ is the dark energy equation of state. From Eq. \ref{eq:bartelmann-kappa} 
we see that if each $\delta(\bm{\theta},z)$ is drawn from a lognormal distribution then 
$\kappa(\bm{\theta},z_{\mr{s}})$ is a sum of (correlated) lognormal variables. However, in contrast 
with Gaussian variables, the sum of lognormal variables is not a lognormal variable itself 
\citep[see Figs. \ref{fig:kappa-pdf-diff}, \ref{fig:2D-pdf-delta-kappa} and \ref{fig:2D-pdf-kappa-kappa} 
  and][]{Fenton60mn}. Following the reasoning presented in the end of Sec. \ref{sec:toy-model}, this 
internal inconsistency might be the cause for the lognormal model failure. 

\subsection{Modelling convergence alone as a lognormal field}

Unfortunately, there is no closed expression for the PDF 
of a sum of lognormal variables; this is still an active field of study and several 
approximating formulas have been proposed \citep{Fenton60mn, Schwartz82mn, Lam07mn, Li11amn}. 
Nevertheless, assuming that the joint probability distribution for $\delta(\bm{\theta},z)$ at 
different $z$ is a multivariate lognormal distribution -- i.e. $\ln[\delta(\bm{\theta},z)]$ are drawn from 
a multivariate Gaussian distribution --, it is possible to compute $\kappa(\bm{\theta},z_{\mr{s}})$'s 
moments using the equations described in Appendix \ref{sec:sum-lognormals}. 

Given that $\langle\delta(\bm{\theta},z)\rangle=0$, we have 
$\langle \kappa(\bm{\theta},z_{\mr{s}})\rangle=0$ as well. The convergence variance 
and skewness are:

\begin{equation}
\mr{Var}[\kappa(z_{\mr{s}})]=\iint_0^{z_{\mr{s}}}
K(z_1,z_{\mr{s}})K(z_2,z_{\mr{s}})\xi_{\delta\delta}(z_1,z_2)\mr{d}z_1\mr{d}z_2,
\label{eq:kappa-var} 
\end{equation}

\begin{equation}
\begin{split}
&\mr{Skew}[\kappa(z_{\mr{s}})]=\frac{1}{\mr{Var}^{3/2}[\kappa(z_{\mr{s}})]}\cdot\\
\iiint_0^{z_{\mr{s}}}&K(z_1,z_{\mr{s}})K(z_2,z_{\mr{s}})K(z_3,z_{\mr{s}})[3\xi_{\delta\delta}(z_1,z_2)\xi_{\delta\delta}(z_2,z_3) + \\
&\xi_{\delta\delta}(z_1,z_2)\xi_{\delta\delta}(z_2,z_3)\xi_{\delta\delta}(z_3,z_1)]\mr{d}z_1\mr{d}z_2\mr{d}z_3,
\end{split}
\label{eq:kappa-skew} 
\end{equation}
respectively, where $\xi_{\delta\delta}(z,z')=\langle \delta(\bm{\theta},z)\delta(\bm{\theta},z')\rangle$ 
is the matter density contrast LoS correlation function. Eq. \ref{eq:kappa-var} 
does not provide any new information since the variance is already fixed by the convergence power spectrum. 
Eq. \ref{eq:kappa-skew}, however, puts a constraint over the convergence distribution's shape; 
if one wants to approximate the convergence as a lognormal variable, it can be used in conjunction with Eq. 
\ref{eq:moments2shift} to specify the distribution's shift parameter $\lambda_i$ directly from theory. 
This is useful since previous methods for determining $\lambda_i$ relied on 
computationally expensive ray tracing through $N$-body simulations \citep[e.g.][]{Taruya02mn, Hilbert11mn}. 
Another alternative method for computing lensing PDFs is given by \citet{Kainulainen11mn}.

To verify these conclusions numerically, we used {\sc flask} to create 12.5 million lognormal realisations of the 
LoS matter density in 41 top-hat redshift bins of width $\Delta z=0.05$ in the range $0.05<z<2.10$ 
and to obtain the convergence at $z=2.10$ for each realisation using Eq. \ref{eq:bartelmann-kappa}. 
We then measured the statistics of the convergence sample and compared with the values expected from 
theory. As Table \ref{tab:intdens2kappa} shows, they all match to 1\% or better. 
Using Eqs. \ref{eq:moments2shift}, \ref{eq:mu-from-moments} and \ref{eq:sigma-from-moments} we 
can compute the parameters of the lognormal distribution that would satisfy such statistics; 
these parameters are shown in the bottom part of Table \ref{tab:intdens2kappa}.  
 
\begin{table}
\centering
\begin{tabular}{lcc}
\hline
Statistic     & Numerical           & Theory  \\
\hline
Mean          & $3.27\times 10^{-6}$ & 0       \\        
Std. Dev.     & 0.02182             & 0.02189 \\ 
Skewness      & 0.513               & 0.508   \\
\hline
Lognormal fit &                 &         \\
\hline
$\mu$         & -2.063              & -2.050  \\ 
$\sigma$      & 0.1682              & 0.1665  \\
$\lambda$     & 0.1288              & 0.1306  \\
\hline
\end{tabular}
\caption{The top part shows the mean, standard deviation and skewness of the convergence distribution 
obtained through density LoS integration (middle column) and the expected values from theory 
(0 and those given by Eqs. \ref{eq:kappa-var} and \ref{eq:kappa-skew}, last column). The bottom part 
shows the lognormal distribution parameters that would reproduce the statistics above.}
\label{tab:intdens2kappa} 
\end{table} 

\begin{figure}
  \includegraphics[width=1\columnwidth]{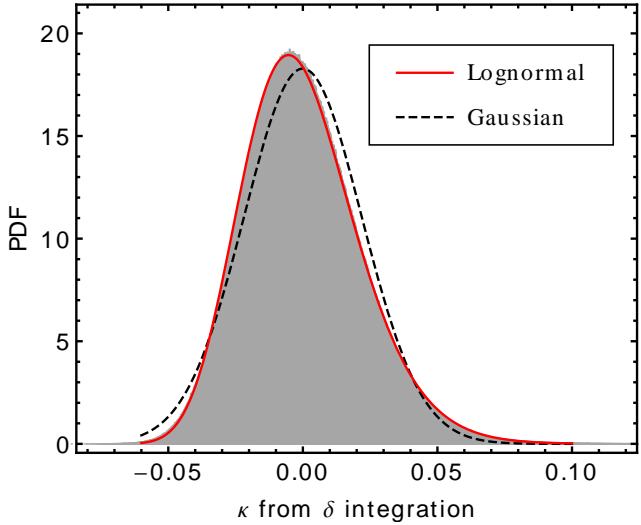}
  \caption{The shaded region is a histogram for the 12.5 million convergences at $z=2.10$
    obtained from lognormal density LoS integration. The black dashed (red solid) 
    line shows a Gaussian (lognormal) distribution that have the 
    same mean and variance (mean, variance and skewness) as the convergence; their parameters 
    are given by the theoretical values in Table \ref{tab:intdens2kappa}. The 
    lognormal model performs much better than the Gaussian but significant deviations exist; 
    these are better seen in Fig. \ref{fig:kappa-pdf-diff}.}
  \label{fig:kappa-hist}
\end{figure}

\begin{figure}
  \includegraphics[width=1\columnwidth]{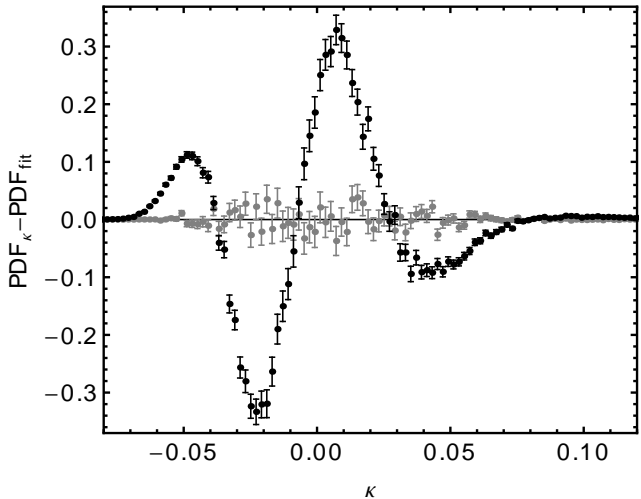}
  \caption{Difference between the PDF of the convergence at $z=2.10$ 
    obtained from lognormal density LoS integration and the lognormal 
    PDF with the same first three moments (black points): significant deviations 
    exist. The grey points is an example of differences one would get from this 
    moment matching technique if the convergence was indeed lognormal: they would be 
    consistent with zero. The error bars represent the Poisson noise inside each bin.}
  \label{fig:kappa-pdf-diff}
\end{figure}

\begin{figure*}
  \includegraphics[width=0.9\textwidth]{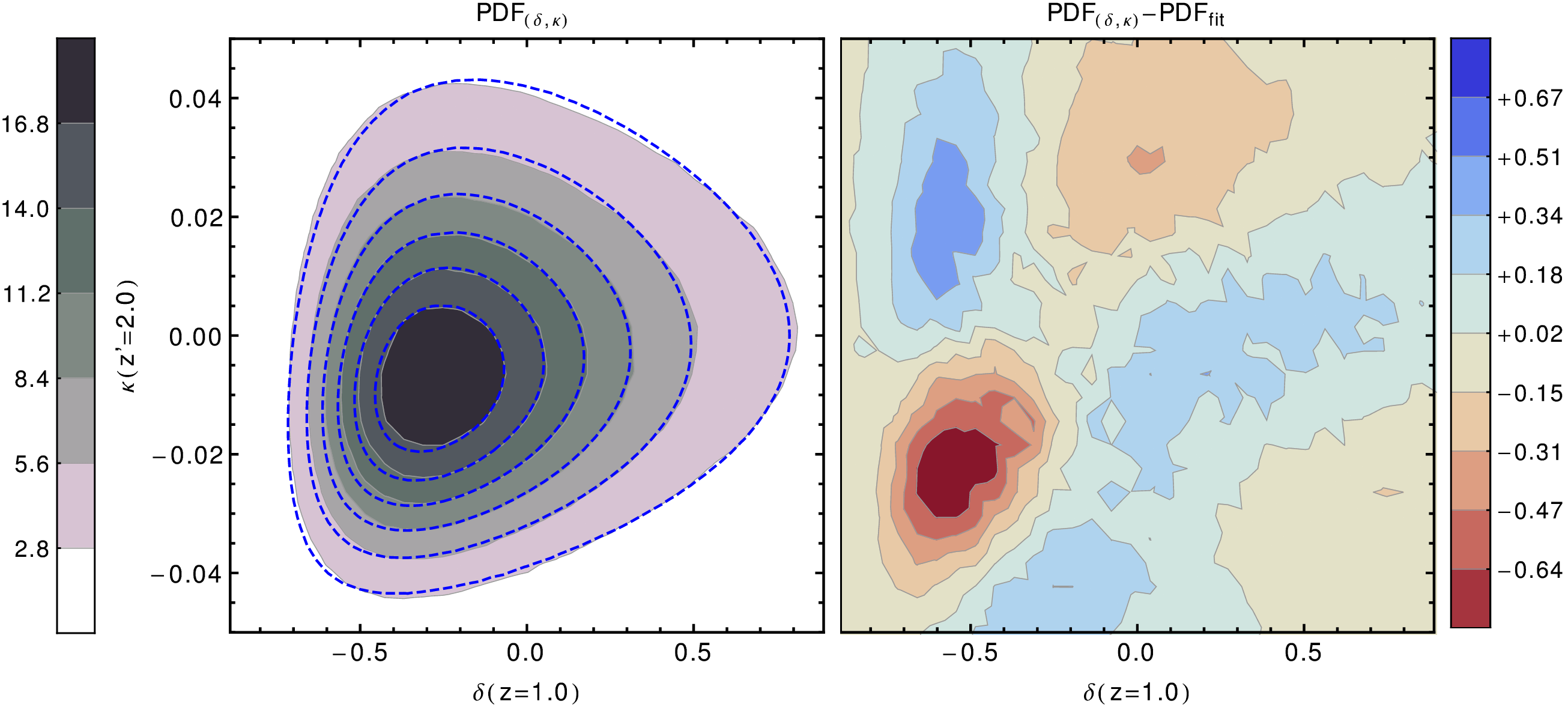}
  \caption{\emph{Left panel:} the shaded regions represent the joint PDF 
    for density contrast $\delta(\bm{\theta},z)$ at redshift $z=1.0$ and convergence 
    $\kappa(\bm{\theta},z')$ at redshift $z'=2.0$, when $\delta$ is drawn from a lognormal distribution 
    and $\kappa$ computed by density LoS integration (darker regions have higher probability densities), estimated using 
    $\sim 12.5\times10^6$ realisations.
    The dashed blue contours represent the two-dimensional multivariate lognormal PDF  
    whose means, covariances and skewnesses are the same as those for the former PDF. If these two PDFs were 
    the same, the contours would overlap. \emph{Right panel:} this contour plot shows the difference between 
    the two PDFs in the left panel (density LoS integration PDF minus lognormal PDF). The red regions have negative 
    values while blue regions have positive values.}
  \label{fig:2D-pdf-delta-kappa}
\end{figure*}

\begin{figure*}
  \includegraphics[width=0.9\textwidth]{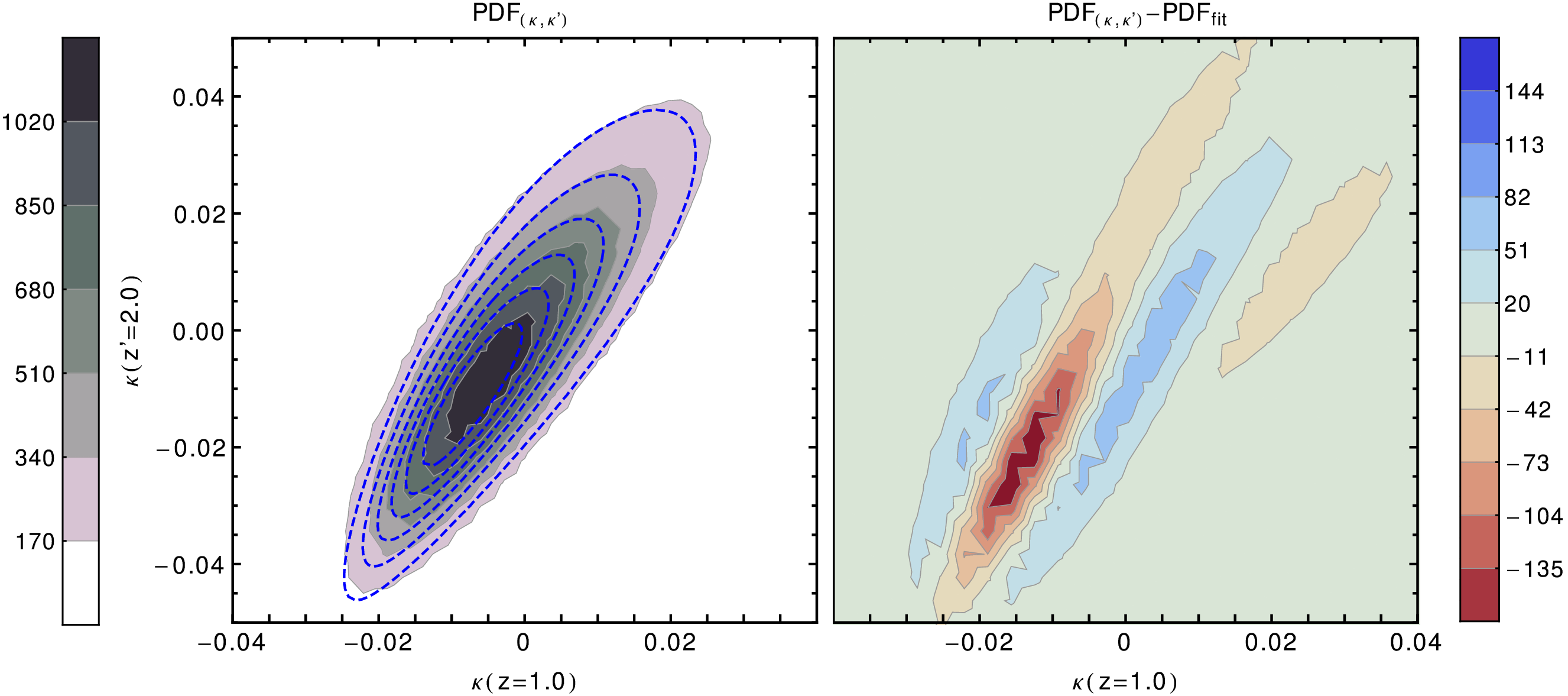}
  \caption{Same as Fig. \ref{fig:2D-pdf-delta-kappa} but for two convergences computed by density LoS integration 
    (along the same LoS) for sources at redshifts $z=1.0$ and $z'=2.0$. 
  }
  \label{fig:2D-pdf-kappa-kappa}
\end{figure*}

Fig. \ref{fig:kappa-hist} shows that the theoretical parameters from Table \ref{tab:intdens2kappa} -- 
chosen to reproduce the first three moments of the distribution -- indeed provide a good fit for 
the convergence derived as a sum of correlated lognormal variables. The reproduction however is not 
perfect as can be seen in Fig. \ref{fig:kappa-pdf-diff}. A similar analysis was performed for 
two-dimensional distributions and the results are presented in Figs. \ref{fig:2D-pdf-delta-kappa} and 
\ref{fig:2D-pdf-kappa-kappa}, where we see that the disagreement between the simulated distribution 
and the lognormal model is larger for the convergence--convergence joint distribution. This is reasonable 
given that, in this case, both one-dimensional marginals were distorted away from the lognormal model. 
Therefore, we might expect that higher-dimensional convergence distributions will be even less well 
approximated by the multivariate lognormal model.

At this point it is important to stress that, since convergence is not strictly a lognormal 
variable, different methods of determining its shift parameter will result in different values. 
One possible method -- which was implemented by \citet{Taruya02mn} -- is to set the shift 
parameter as the minimum attainable convergence; in a lognormal density model, this is clearly 
the convergence for the empty line of sight [Eq. \ref{eq:bartelmann-kappa} with 
$\delta(\bm{\theta},z)=-1$] which is hard to be extracted from simulations (or observations) 
given that these consist of finite samples. Another method \citep[implemented by][]{Hilbert11mn} is 
to perform a least-squares fit to the convergence PDF. Table \ref{tab:shift-methods} compares 
the values obtained from each method when applied to the convergence modelled as a sum of lognormal 
variables (density LoS integration) and as a lognormal variable itself. While 
for the lognormal case the methods agree and return the shift specified \emph{a priori} 
(apart from the minimum value method which suffers from finite sampling), there is no 
agreement for the other case (which unfortunately is more realistic). This means that the 
lognormal approximation for the convergence field cannot reproduce every aspect of the distribution 
and choices have to be made: if one wants to reproduce the convergence skewness, the moment 
matching method should be used; if one wants to reproduce the PDF shape as close as possible, 
the least-squares method should be preferred; and so on.

\begin{table}
\centering
\begin{tabular}{lcc}
\hline
Method                  & Sum of lognormals & Lognormal  \\
\hline
Empty LoS               &  0.2115           &   --       \\
True value              &    --             & 0.1300     \\
Moment matching         &  0.1288           & 0.1306     \\
PDF least square        &  0.1482           & 0.1310     \\
Min. value              &  0.0858           & 0.0779     \\
\hline
\end{tabular}
\caption{Possible methods of determining the shift parameter of a lognormal distribution 
fit to a convergence sample: matching the first three moments; using the least squares 
method as in \citet{Hilbert11mn}; selecting the minimum value from the sample as in 
\citet{Taruya02mn}. These are applied to a sample of convergences that are sums of 
correlated lognormal densities (middle column) and to a sample of true lognormal convergences 
(last column). The results are compared to the true value assigned to the distribution 
(no such thing for the sum of lognormals) and to the empty line-of-sight value. Comparing 
the latter with the lognormal sample does not make sense as the latter had the shift parameter 
set \emph{ad hoc} to match the value obtained for the density LoS integration 
under the moment matching method.}
\label{tab:shift-methods}
\end{table}

\subsection{Quantifying the deviation from lognormal distribution}
\label{sec:pdf-fit}

To better describe the shape of the convergence 1D marginal distributions obtained 
by lognormal density LoS integration we used the minimum $\chi^2$ method to fit the 
following formula to that PDF:

\begin{equation}
f_{\mr{ABC}}(\kappa) = \frac{1}{\sqrt{2\pi}s}\exp\left\{-\frac{[ABC'(\kappa)-m]^2}{2s^2}\right\}\frac{\mr{d}ABC'}{\mr{d}\kappa},
\label{eq:abc-pdf}
\end{equation}
\begin{equation}
ABC'(\kappa) = \frac{1}{t}\sinh\left\{\frac{t\kappa_0}{\gamma}\left[\left(\frac{\kappa}{\kappa_0} + 1\right)^\gamma - 1\right]\right\}.
\label{eq:abc-trafo}
\end{equation}
In the equations above, $ABC'$ is a slightly modified version of the $ABC$ Gaussianization transformation 
\citep[that is, it transforms variables that follow more general distributions into Gaussian ones; ][]{Schuhmann15mn} 
when its parameters $t$ and $\gamma$ are restricted to $t>0$ and $\gamma\neq 0$.\footnote{In reality, the $f_{\mr{ABC}}(\kappa)$ 
fitting was performed with unrestricted $t$ and $\gamma$, but the best fit remained in the $t>0$ and $\gamma\neq 0$ region. 
Given that the $ABC$ transformation is a piecewise function, we only show here the $t>0$ and $\gamma\neq 0$ sub-function.} 
The more general $\kappa$ PDF and the PDF of a Gaussian variable $z$ are related by a simple change of variables: 
$f(\kappa)\mr{d}\kappa = G(z)\mr{d}z$, where $G(z)$ is a Gaussian PDF.

The simulated data used were $\sim 3.1$ million convergences for sources at each of the redshifts 
$z$ specified in Table \ref{tab:pdf-fit-pars}, convolved with Gaussian window 
functions of radius (standard deviation) 1.23 arcmin. These were produced by integrating the 
lognormal density simulated in 40 equal-width redshifts bins in the range $0.025<z<2.025$ (details 
of this procedure are given in Sec. \ref{sec:solutions}). For each redshift, the convergences were distributed into 500 bins 
covering the full data range, but when fitting the function given by Eq. \ref{eq:abc-pdf} we restricted 
our analysis to the range $\kappa_{\mr{min}}<\kappa<\kappa_{\mr{max}}$ that does not include bins with zero counts. 
Together with this range, Table \ref{tab:pdf-fit-pars} presents for each redshift the $f_{\mr{ABC}}(\kappa)$ 
parameters $m$, $s$, $t$, $\gamma$  and $\kappa_0$ that best fit the convergence PDFs, along with the 
best-fit $p$-value. In most cases, the $p$-values indicate that the fits are quite good. 

\begin{table*}
\centering
\begin{tabular}{ccccccccc}
\hline
$z$ & $\kappa_{\mr{min}}$ & $\kappa_{\mr{max}}$ & $m$ & $s$ & $t$ & $\gamma$ & $\kappa_0$ & $p$-value\\ 
\hline
0.525 & -0.0164 & 0.105 & -0.001458 & 0.00690 & 31.034 & -1.383 & 0.040 & 0.002 \\
0.575 & -0.0191 & 0.121 & -0.001516 & 0.00764 & 27.047 & -1.663 & 0.053 & 0.103 \\
0.625 & -0.0213 & 0.117 & -0.001566 & 0.00837 & 23.974 & -1.998 & 0.069 & 0.052 \\
0.675 & -0.0234 & 0.111 & -0.001609 & 0.00907 & 21.405 & -2.400 & 0.089 & 0.281 \\
0.725 & -0.0257 & 0.123 & -0.001647 & 0.00976 & 19.019 & -2.748 & 0.111 & 0.425 \\
0.775 & -0.0286 & 0.132 & -0.001680 & 0.01044 & 17.368 & -3.346 & 0.145 & 0.385 \\
0.825 & -0.0315 & 0.125 & -0.001708 & 0.01109 & 15.864 & -4.070 & 0.188 & 0.231 \\
0.875 & -0.0342 & 0.141 & -0.001734 & 0.01173 & 14.419 & -4.766 & 0.235 & 0.270 \\
0.925 & -0.0361 & 0.139 & -0.001755 & 0.01235 & 13.412 & -6.209 & 0.322 & 0.648 \\
0.975 & -0.0382 & 0.137 & -0.001772 & 0.01295 & 12.596 & -8.798 & 0.476 & 0.715 \\
1.025 & -0.0409 & 0.147 & -0.001786 & 0.01354 & 12.060 & -16.499 & 0.922 & 0.074 \\
1.075 & -0.0435 & 0.154 & -0.001799 & 0.01411 & 11.249 & -30.820 & 1.807 & 0.389 \\
1.125 & -0.0461 & 0.141 & -0.001812 & 0.01466 & 10.055 & -30.820 & 1.935 & 0.172 \\
1.175 & -0.0484 & 0.144 & -0.001823 & 0.01520 & 9.048 & -30.820 & 2.064 & 0.859 \\
1.225 & -0.0501 & 0.151 & -0.001832 & 0.01573 & 8.194 & -30.820 & 2.196 & 0.161 \\
1.275 & -0.0517 & 0.152 & -0.001841 & 0.01624 & 7.490 & -30.820 & 2.324 & 0.399 \\
1.325 & -0.0535 & 0.151 & -0.001846 & 0.01673 & 6.670 & -30.820 & 2.460 & 0.356 \\
1.375 & -0.0560 & 0.153 & -0.001851 & 0.01722 & 6.045 & -30.820 & 2.594 & 0.068 \\
1.425 & -0.0584 & 0.172 & -0.001857 & 0.01769 & 5.447 & -30.820 & 2.726 & 0.098 \\
1.475 & -0.0603 & 0.171 & -0.001859 & 0.01814 & 4.913 & -30.820 & 2.861 & 0.263 \\
1.525 & -0.0617 & 0.161 & -0.001862 & 0.01859 & 4.426 & -30.820 & 2.998 & 0.060 \\
1.575 & -0.0637 & 0.173 & -0.001863 & 0.01902 & 3.955 & -30.820 & 3.133 & 0.012 \\
1.625 & -0.0652 & 0.155 & -0.001864 & 0.01945 & 3.642 & -30.820 & 3.272 & 0.001 \\
1.675 & -0.0663 & 0.181 & -0.001864 & 0.01986 & 3.103 & -30.820 & 3.407 & 0.050 \\
1.725 & -0.0672 & 0.186 & -0.001865 & 0.02027 & 2.721 & -30.820 & 3.543 & $3.7\times 10^{-4}$ \\
1.775 & -0.0685 & 0.169 & -0.001864 & 0.02066 & 2.393 & -30.820 & 3.681 & 0.003 \\
1.825 & -0.0704 & 0.178 & -0.001864 & 0.02105 & 2.124 & -30.820 & 3.818 & 0.001 \\
1.875 & -0.0725 & 0.179 & -0.001863 & 0.02143 & 1.798 & -30.820 & 3.954 & $2.2\times 10^{-4}$ \\
1.925 & -0.0747 & 0.166 & -0.001862 & 0.02180 & 1.416 & -30.820 & 4.088 & 0.017 \\
1.975 & -0.0768 & 0.191 & -0.001862 & 0.02216 & 0.972 & -30.820 & 4.221 & 0.058 \\
2.025 & -0.0788 & 0.187 & -0.001860 & 0.02251 & 0.509 & -30.820 & 4.358 & 0.022 \\
\hline
\end{tabular}
\caption{Fit to the marginal distribution of the convergence obtained by LoS integration of the lognormal density. 
  The columns are, from left to right: the sources' redshift $z$, the minimum and maximum convergences 
  used in the fit, the five $f_{\mr{ABC}}(\kappa)$ parameters and the fit $p$-value. From $z=1.075$ onwards, 
  $\gamma$ is fixed to the best-fitting parameter for that redshift to avoid numerical instabilities.}
\label{tab:pdf-fit-pars}
\end{table*}

\begin{figure}
  \includegraphics[width=1.0\columnwidth]{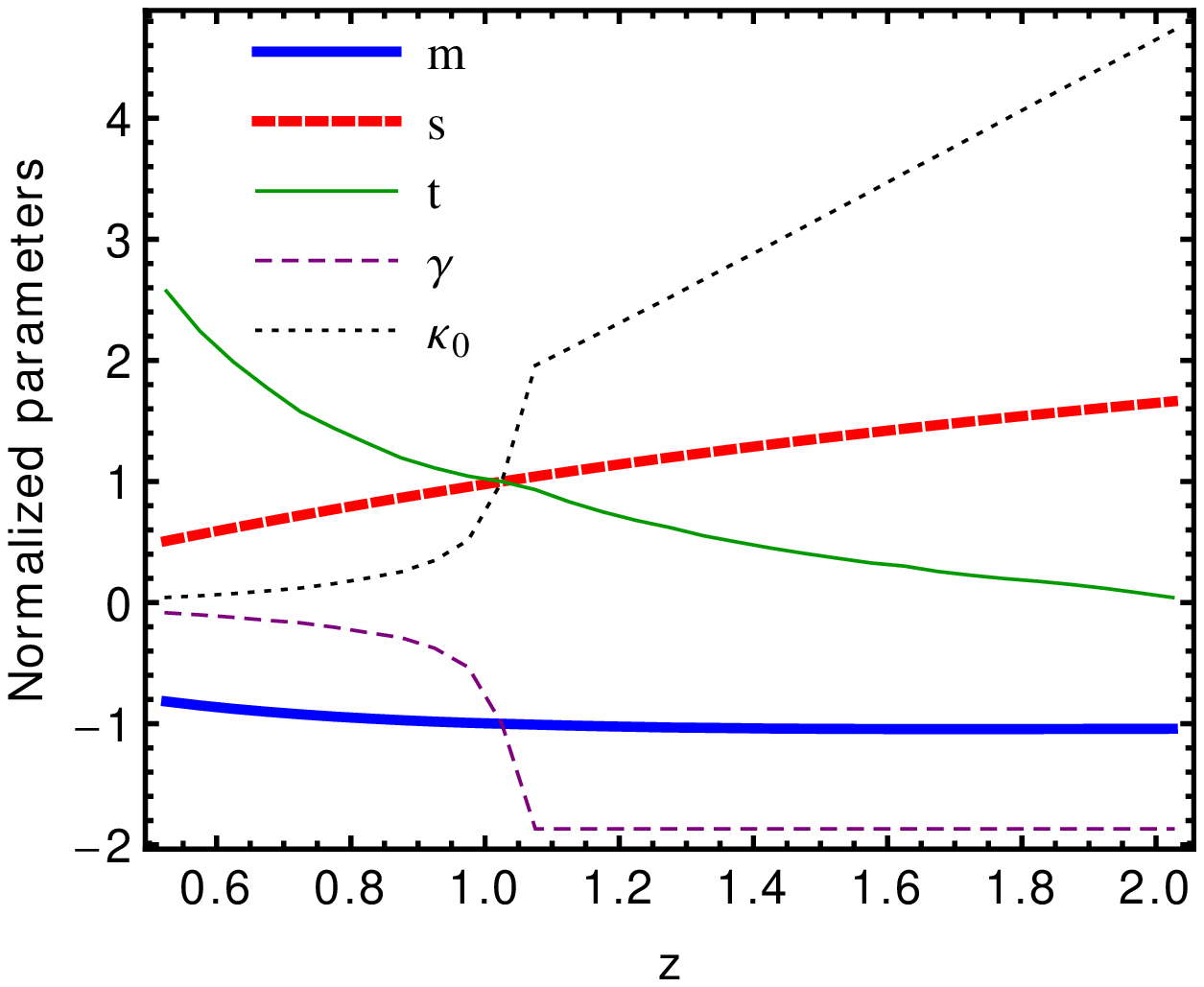}
  \caption{Best-fitting parameters from Table \ref{tab:pdf-fit-pars}, normalised by their absolute values at 
    $z=1.025$.}
  \label{fig:abc-fit-pars}
\end{figure}

\begin{figure}
  \includegraphics[width=1.0\columnwidth]{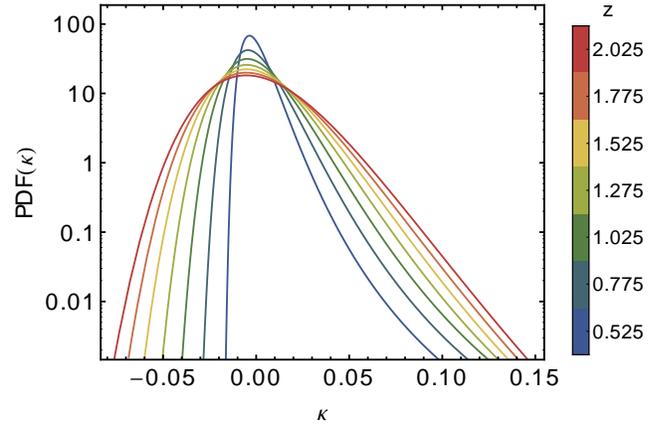}
  \caption{Best-fitting $f_{\mr{ABC}}(\kappa)$ distributions for the convergence obtained from lognormal density 
    LoS integration, for various redshifts. As the redshift of the sources increases, the distribution gets closer 
    to a Gaussian.}
  \label{fig:abc-pdfs}
\end{figure}

As the redshift increases, the number of density bins that get summed into the convergence increases and 
its distribution gets closer to a Gaussian due to the central limit theorem, thus making the distribution 
less complex and requiring less parameters. This is manifested by the strong correlation between $\gamma$ and 
$\kappa_0$ that grows with redshift up to a point where they diverge and the fits get unstable. 
To avoid this issue, for redshift 1.075 onwards we fixed $\gamma$ at the best-fitting value obtained at that 
redshift. The parameters from Table \ref{tab:pdf-fit-pars} are also presented in Fig. \ref{fig:abc-fit-pars}. 
The resulting $f_{\mr{ABC}}(\kappa)$ for various redshifts are shown in Fig. \ref{fig:abc-pdfs}.

\section{The line-of-sight integration solution}
\label{sec:solutions}

The limitations for simulating correlated density and convergence presented in Sec. 
\ref{sec:distortions} can be circumvented in three ways. First, by simulating Gaussian instead of 
lognormal fields; as explained in Sec. \ref{sec:toy-model}, Gaussian variables are less 
limited in terms of valid covariance matrices than lognormal ones (as a trade-off, however, one loses 
the skewness and minimum boundary of the lognormal distribution). A second option is to distort the 
input power spectra so they produce positive-definite covariance matrices; once you have a valid 
covariance matrix for the associated Gaussian multipoles, the lognormal simulation can proceed 
without further issues. This is acceptable if the application intended for the simulations does 
not require the input $C(\ell)$s to be linked to a particular cosmological model or if the fractional 
changes applied to the input $C(\ell)$s are within the precision tolerance. The third option is 
to generate an only-density lognormal simulation and obtain the convergence by performing an 
approximated LoS integration through a weighted Riemann sum of the simulated densities 
in the redshift bins; as presented in Sec. \ref{sec:distortions}, density realisations are 
not plagued by lognormal limitations.

As shown in Figs. \ref{fig:kappa-hist} to \ref{fig:2D-pdf-kappa-kappa}, such integration leads to 
a convergence field that follows a distribution different from the lognormal (although fairly 
similar). To test if such a convergence field follows the expected statistics, we created 
1000 full-sky simulations of the density field in 40 contiguous redshift bins of width 
$\Delta z = 0.05$, spanning the range $0.025<z<2.025$, and computed the convergence by approximating 
the integral in Eq. \ref{eq:bartelmann-kappa} by a Riemann sum. We approximated the continuous 
density contrast $\delta(\bm{\theta},z)$ by its  average inside redshift bins 
$\bar{\delta}(\bm{\theta},z_i)$ (which already is the {\sc class} output) and the kernel 
$K(z,z_{\mr{s}})$ by its average inside the same bins $\bar{K}(z_i,z_{\mr{s}})$:

\begin{equation}
\kappa(\bm{\theta},z_{\mr{s}}) \simeq \sum_i \bar{K}(z_i,z_{\mr{s}}) \bar{\delta}(\bm{\theta},z_i)\Delta z_i.
\label{eq:kappa-riemann-sum}
\end{equation}
We then recovered the power spectra from the convergence field computed as above and compared with the 
{\sc class} output. It is worth noting that the effects of such approximation can be predicted from 
$C^{\bar{\delta}(z)\bar{\delta}(z')}(\ell)$, the spectra for the average density contrasts $\bar{\delta}$ at 
redshift bins centred at $z$ and $z'$, computed by {\sc class}. The power spectra expected for 
convergence $\kappa$ at redshifts $z_{\mr{s}}$ and $z'_{\mr{s}}$, and for density contrast $\bar{\delta}$ 
at redshift $z$ and convergence at redshift $z_{\mr{s}}$, are: 

\begin{equation}
\begin{split}
&\tilde{C}^{\kappa(z_{\mr{s}})\kappa(z'_{\mr{s}})}(\ell) = \\
&\sum_i\sum_j \bar{K}(z_i,z_{\mr{s}}) \bar{K}(z_j,z'_{\mr{s}}) C^{\bar{\delta}(z_i)\bar{\delta}(z_j)}(\ell) \Delta z_i\Delta z_j,
\end{split}
\label{eq:clkk-riemann-sum}
\end{equation}

\begin{equation}
\tilde{C}^{\bar{\delta}(z)\kappa(z_{\mr{s}})}(\ell) =
\sum_i\bar{K}(z_i,z_{\mr{s}})C^{\bar{\delta}(z)\bar{\delta}(z_i)}(\ell) \Delta z_i.
\label{eq:cldk-riemann-sum}
\end{equation}

\begin{figure*}
  \includegraphics[width=1\textwidth]{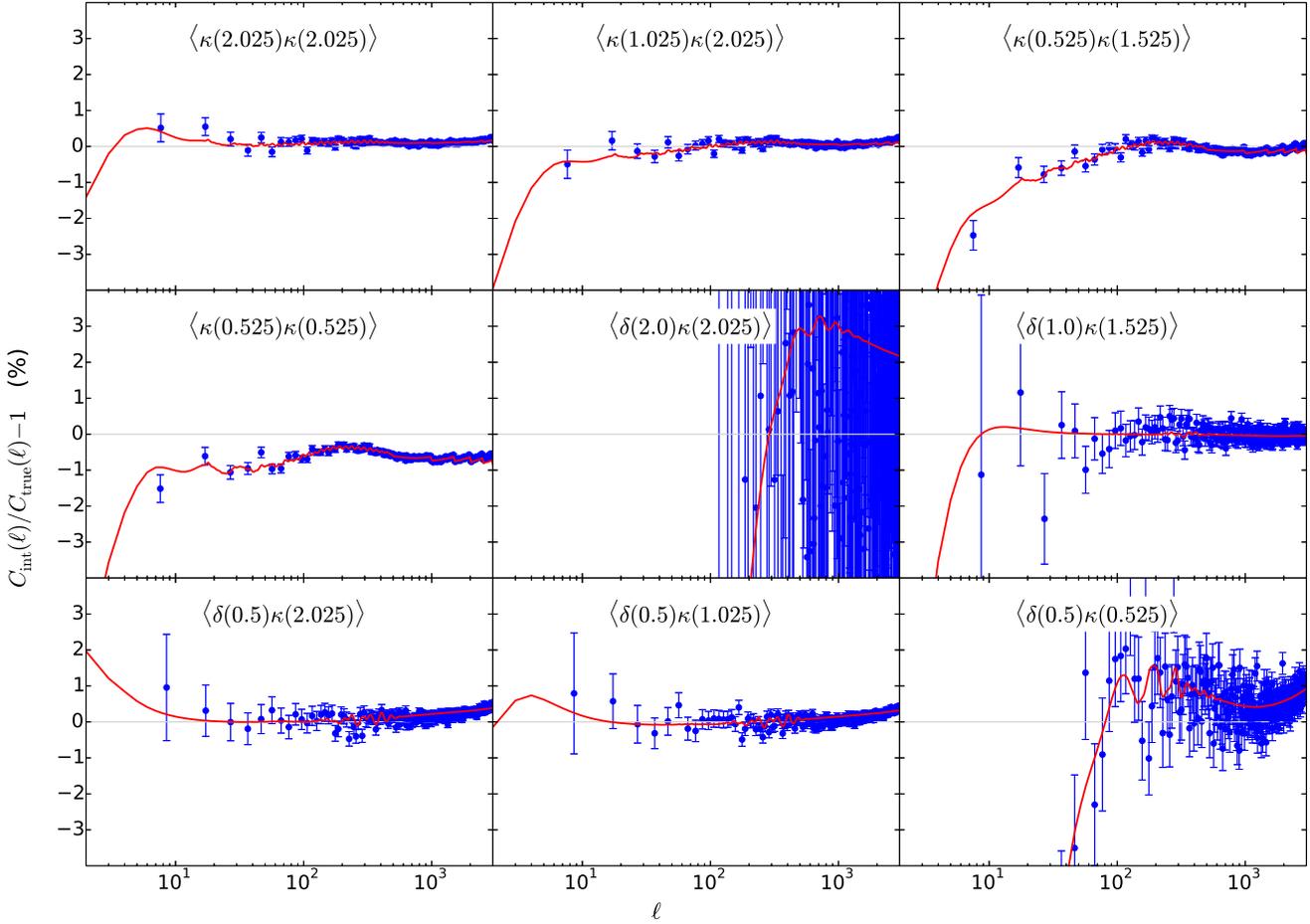}
  \caption{Fractional difference between the angular power spectra for the convergence computed as 
    a LoS Riemann sum of the simulated density, $C_{\mr{int}}(\ell)$, and the one computed by {\sc class}, 
    $C_{\mr{true}}(\ell)$. The red curves show the theoretical expectation of the results, given by 
    Eqs. \ref{eq:clkk-riemann-sum} and \ref{eq:cldk-riemann-sum}, and the blue data points show 
    the average of 1000 power spectra recovered 
    from independent density field realisations, averaged inside 10-$\ell$ bins. The error bars show 
    the error on the mean. The first four subplots show 
    the results for convergence--convergence power spectra while the last five subplots show the 
    results for the density--convergence power spectra. The density was simulated in 40 redshift bins 
    of width $\Delta z = 0.05$, and results were analysed for the bins centred at 0.5, 1.0, 1.5 and 2.0. 
    The convergence was computed for sources at redshifts 0.525, 1.025, 1.525 and 2.025.
  }
  \label{fig:los-int-recov-Cls}
\end{figure*}

A representative set of the results of this comparison is shown in Fig. \ref{fig:los-int-recov-Cls}. All analysed 
power spectra in the range $0.5<z<2.025$ and for $\ell \ga 100$ succeed in reproducing the theoretical 
ones computed by {\sc class} with a 3\% precision (we did not study the spectra for which the 
density is at higher redshift than the convergence since these are very small); in fact, the 
convergence--convergence power spectra in this redshift range all agree to {\sc class} 
$C^{\kappa\kappa}(\ell)$s at 1\% all the way down to $\ell\sim 20$ or better.

One thing that can be seen in Fig. \ref{fig:los-int-recov-Cls} is that the agreement is 
worse at lower redshifts. This happens for two reasons: first, the number of density bins 
used to compute the convergence is smaller (10 for $z=0.5$ against 40 for $z=2.0$), rendering 
a coarser integral approximation; secondly, the truncation of the integral at $z=0.025$ instead 
of at $z=0$ ({\sc class} could not compute power spectra for bins centred at $z<0.05$) is more 
relevant for lower redshifts, thus producing a systematic power loss. We can also see that 
the agreement is worse for lower multipoles. This happens because $C^{\delta(z)\delta(z')}(\ell)$ 
is more sharply peaked in this case and therefore the approximation of the integral by a sum 
is less accurate.

Another aspect worth noting in Fig. \ref{fig:los-int-recov-Cls} is that the density--convergence 
power spectra are well recovered with a precision better that 1\% down to $\ell\sim 10$ in most 
cases, the exception being when the density redshift bin is very close to the convergence 
redshift. This might not be an issue for high redshifts where, as shown by the large error bars, 
such measurements are quite difficult to be performed, but at lower redshifts it can present up to 
3\% deviations at $\ell\ga 60$ and even larger ones at lower multipoles. Part of the problem 
can be accounted by the precision of {\sc class} $C(\ell)$s: the power spectra 
$C^{\bar{\delta}(z)\bar{\delta}(z')}(\ell)$ used to simulate the density fields have to be well tuned with the convergence 
spectra used as $C_{\mr{true}}(\ell)$. Moreover, the outcome of Eq. \ref{eq:cldk-riemann-sum} refers to 
the average density inside a top-hat bin and the convergence at an exact redshift $z_{\mr{s}}$, something 
that cannot be computed by {\sc class}; the comparison was made with the convergence in a top-hat 
redshift bin of width $\Delta z=0.002$ centred at the border of the last density bin used in the integration. 

Finally, Fig. \ref{fig:los-int-recov-Cls} shows that the theoretical prediction from Eqs. 
\ref{eq:clkk-riemann-sum} and \ref{eq:cldk-riemann-sum}, depicted by the red lines, works excellently. 
Thus, if the intended application for the simulations requires convergence fields that accurately 
follow a fiducial $C(\ell)$ but otherwise can deviate from a specific cosmological model, this method 
can be very powerful since comparisons can be made to these predictions.   

\section{FLASK code description}
\label{sec:code-description}

\subsection{Overview}
\label{sec:code-overview}

The purpose of {\sc flask} is to generate two- or three-dimensional random realisations of astrophysical fields 
such as matter, arbitrary tracer densities, weak lensing convergence and shear in a correlated way, 
that is, all simulated fields (e.g. multiple tracers and weak lensing) are connected through the same 
realisation and therefore follow the expected internal cross-correlations provided as input.
 According to the user's choice, the realisations can follow either a multivariate Gaussian distribution 
or a multivariate lognormal distribution in which case each field's one-dimensional marginal distribution is 
lognormal (note that mixed Gaussian and lognormal marginals can also be generated since the Gaussian case 
can be described as a special lognormal case when the field's skewness is zero). In comparison to the 
Gaussian, the lognormal distribution is a better approximation to matter and tracer densities and to the 
lensing convergence; it also does not lead to nonphysical values such as negative densities. 

\begin{figure}
  \includegraphics[width=1\columnwidth,trim={0 4cm 0 0},clip]{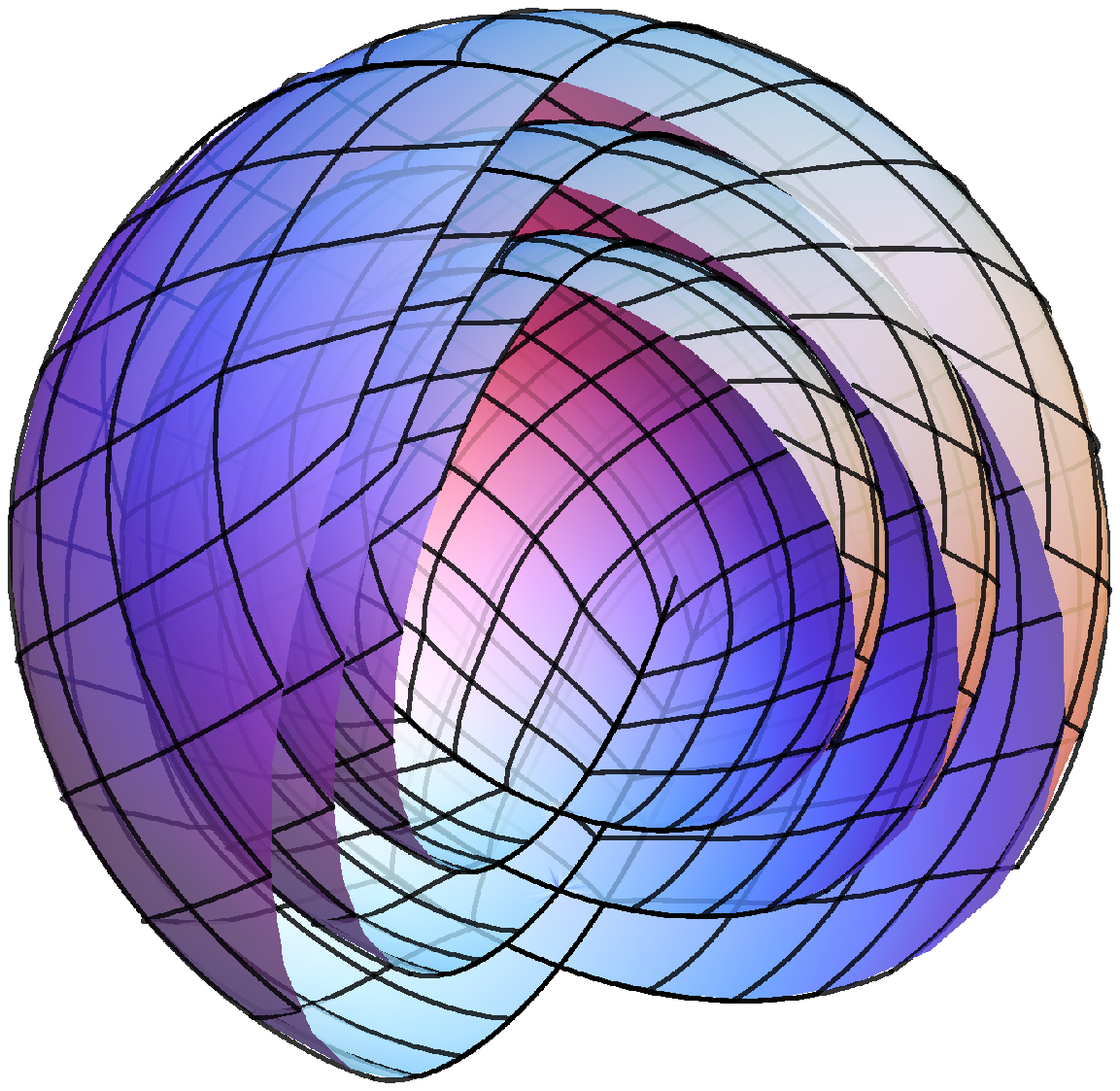}
  \caption{Example of discretization of space used in {\sc flask} (a quarter of the concentric spherical 
    surfaces were removed to ease visualisation). The observer is located in the centre of the spheres. 
    The surfaces represent the cells' boundaries in the radial direction and the black lines represent their 
    angular boundaries. In this example there are two radial slices, 
    each with 192 cells of same angular size. The radial slices can have arbitrary thickness while the angular 
    part in all slices follows the same {\sc healpix} pixelization scheme.}
  \label{fig:onion} 
\end{figure}

Another {\sc flask} feature is that the realisations are created on the full sky using spherical geometry: 
the observer is positioned in the centre of the simulation and the universe is discretized along the LoS 
into spherical shells around the observer of arbitrary thickness (like an onion) with the slices being themselves 
discretized into a fixed number of aligned pixels (see Fig. \ref{fig:onion}). Such geometry allows for 
easy implementation of effects such as evolution with redshift, redshift space distortions and survey 
selection functions. Moreover, it is well matched for upcoming large area surveys than box-shaped simulations. 

The goal of {\sc flask} is to create the full-sky lognormal simulations quickly. Its power spectrum 
realisation approach and its implementation in {\sc C++} using {\sc OpenMP} allows it to generate, 
for instance, 40 full-sky redshift slices of correlated convergence and density fields, each with 
$\sim 50$ million pixels ($N_{\mr{side}}=2048$, which permits analysis to be made up to multipoles 
around $\ell \simeq 4000$), in 10 minutes using a 16 core computer. This redshift and angular resolution 
suffices to create full-size mock catalogues for photometric large-scale structure surveys, such as 
those planned by \emph{Euclid} and LSST, with cosmological signals known to the per-cent level. Another 
relevant aspect of the approach adopted by {\sc flask} is that the statistical properties of the fields 
(i.e. their power spectra and distributions) are defined \emph{a priori} and obeyed by 
construction apart from discretization and truncation inaccuracies that vanish in the limit of 
infinite resolution. This can be an important advantage for certain applications like 
verifying power spectrum and correlation function estimators, evaluating their covariance matrices 
and testing the effects of systematics and statistical fluctuations on these measurements. On the 
down side, the fact that all statistical properties are set by the multivariate lognormal model 
means that the code cannot produce anything different from that (although there is the option to 
model the convergence as a sum of correlated lognormal variables): three-point 
functions, for instance, are bound to behave according to the model and might not give a realistic 
representation of the data; for such applications one might need $N$-body simulations 
\citep[e.g. following][]{Fosalba08mn}. 
    
After generating the fields [which might already include redshift space distortions, magnification 
bias and intrinsic alignments if these were included in the input power spectra -- see 
\citet{Donnacha10mn,Challinor11mn,Blas11mn, Dio13mn}], {\sc flask} can: 
apply survey selection functions (that can be different for different tracers and can be separable 
or not into radial and angular parts), Poisson sample tracers from their density fields, compute 
shear and ellipticities and introduce Gaussian noise in the latter. The final results can be 
output in the form of a source catalogue. Appendix \ref{sec:flask-example} presents an example of 
{\sc flask} usage and output. 

\subsection{Details}
\label{sec:code-details}

\subsubsection{Input}
\label{sec:code-input}

{\sc Flask} is run by calling it on a terminal followed by a configuration file containing all specifications 
needed -- which can also be overridden through the command line. Besides a keyword setting the 
type of simulation to be performed -- Gaussian, lognormal or homogeneous (i.e. Poisson sampling from density 
fields containing no structure) --, two inputs described in 
the configuration file fully specify all statistical properties of the fields: a file containing 
a table of fields' means, shifts and redshift ranges and a set of angular auto and cross power 
spectra $C_{\mr{ln}}^{ij}(\ell)$ for all fields at all redshift slices that must be provided by the user 
(the indices $i$ and $j$ cycle both through fields and redshift slices). 
These $C_{\mr{ln}}^{ij}(\ell)$s can be calculated by public codes such as {\sc class} 
\citep{Blas11mn, Dio13mn}, by {\sc camb sources} \citep{Challinor11mn} or by any other routine. 
In order to fix the fields' properties, all cross-correlations have to be specified [there is an option to treat missing 
$C_{\mr{ln}}^{ij}(\ell)$s as zero, that is, the field/redshift $i$ is uncorrelated with the 
field/redshift $j$]. For instance, $N_{\mr{f}}$ fields described in $N_{\mr{z}}$ redshift 
slices require a total of $N_{\mr{f}}N_{\mr{z}}(N_{\mr{f}}N_{\mr{z}}+1)/2$ $C_{\mr{ln}}^{ij}(\ell)$s to be 
fully specified. Each field can be simulated in a different number of redshift slices which can 
have different ranges as well.

\subsubsection{Obtaining the associated Gaussian power spectra}
\label{sec:ln2g-Cl}

The process of simulating a lognormal field involves first generating a Gaussian one and 
exponentiating it afterwards. To associate the statistical properties of the Gaussian to the 
lognormal field using Eq. \ref{eq:gaussian-cov} we assume statistical homogeneity and isotropy and 
that the field value at each point in space is a random variable. Since Eq. \ref{eq:gaussian-cov} is 
local, the correlation functions of the lognormal and Gaussian fields $\xi_{\mr{ln}}^{ij}(\theta)$ and 
$\xi^{\mr{g}}_{ij}(\theta)$ have the same form as that equation:

\begin{equation}
\xi_{\mr{g}}^{ij}(\theta)= \ln\left[\frac{\xi_{\mr{ln}}^{ij}(\theta)}{\alpha_i \alpha_j}+1\right].
\label{eq:xi-relation}
\end{equation}

Even though $\xi_{\mr{g}}^{ij}(\theta)$ specify a covariance matrix for the field 
in all points in space that could be used to generate correlated Gaussian variables, 
in practice this approach is impossible due to its size. A more economical approach is to 
go to harmonic space since isotropy leads to independent multipoles. 
The relations between the correlation function $\xi^{ij}(\theta)$ and the power spectrum $C^{ij}(\ell)$ 
are given by Eqs. \ref{eq:legendre-trafo-cl} and \ref{eq:legendre-trafo-xi}. 
To obtain the angular power spectra $C_{\mr{g}}^{ij}(\ell)$ for the Gaussian fields we: 
(i) transform the input $C_{\mr{ln}}^{ij}(\ell)$ to real space using Eq. \ref{eq:legendre-trafo-xi}; 
(ii) compute $\xi_{\mr{g}}^{ij}(\theta)$ using Eq. \ref{eq:xi-relation}; and (iii) go back to harmonic space with 
Eq. \ref{eq:legendre-trafo-cl}. In practice the transformations to and from harmonic space are performed 
using the discrete Legendre transform routines implemented in 
{\sc s2kit}\footnote{\texttt{\footnotesize{http://www.cs.dartmouth.edu/\char`~geelong/sphere}}} \citep{Kostelec00mn}. 
For Gaussian realisations, this transformation is skipped and the input power spectra are directly 
used to generate the multipoles; in other words, in this case {\sc flask} simply sets 
$C_{\mr{g}}^{ij}(\ell)=C_{\mr{ln}}^{ij}(\ell)$.

\begin{figure}
  \includegraphics[width=1\columnwidth]{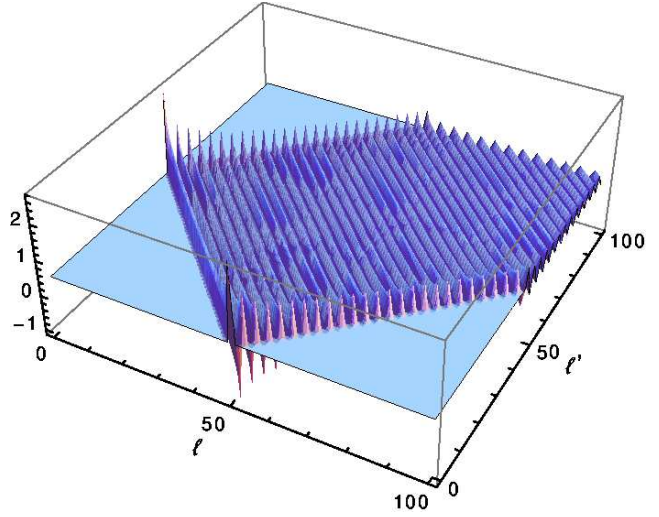}
  \caption{Value of the first Wigner 3-$j$ symbol in Eq. \ref{eq:g-multipole-convol} for $L=50$. 
This serves as an indication that Gaussian multipoles at $l>L$ contribute to the lognormal multipole. 
The structure shown is similar for higher $L$ as well.}
\label{fig:3j-symbol}  
\end{figure}

\begin{figure}
  \includegraphics[width=1\columnwidth]{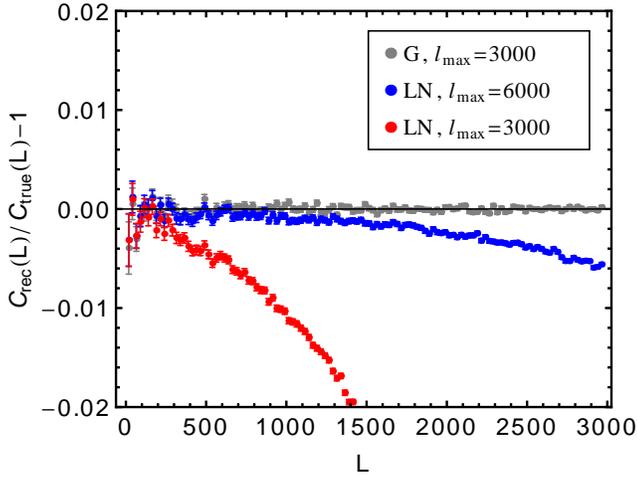}
  \caption{Fractional difference between true angular power spectrum and the average of recoveries from 
    400 full-sky realisations, averaged inside 25-$\ell$ bins. The data points represent: 
    Gaussian realisation with bandlimit $l_{\mr{max}}=3000$ (grey) 
    and lognormal realisations with associated Gaussian field bandlimits $l_{\mr{max}}=6000$ (blue) and 
    $l_{\mr{max}}=3000$ (red). In this example, the simulation of the Gaussian field up to $l_{\mr{max}}=6000$ 
    is enough to get a precision of less than one per cent for the lognormal field power spectrum up to 
    $L=3000$. The error bars represent the error on the mean.}
  \label{fig:ln-power-loss}  
\end{figure}

It is interesting and important to note that since the relation between lognormal and Gaussian 
fields $X_i(\hat{\bm{\theta}})$ and $Z_i(\hat{\bm{\theta}})$ is local in real space, it is non-local 
in harmonic space, i.e. each multipole of the lognormal field depends on a mix of the Gaussian 
multipoles. This can be demonstrated through a series expansion of the exponential:

\begin{equation}
X_i(\hat{\bm{\theta}}) = e^{Z_i(\hat{\bm{\theta}})}-\lambda_i 
\simeq 1-\lambda_i+Z_i(\hat{\bm{\theta}})+\frac{Z_i^2(\hat{\bm{\theta}})}{2}+\cdots.
\label{eq:exp-series}
\end{equation} 
We can expand $X_i(\hat{\bm{\theta}})$ and $Z_i(\hat{\bm{\theta}})$ in spherical harmonics:

\begin{equation}
a(\hat{\bm{\theta}})=\sum_{l,m}a_{lm}Y_{lm}(\hat{\bm{\theta}})
\label{eq:alm-expansion}
\end{equation}
and show that the contribution $X_{i,LM}^{\mr{(2)}}$ of the last written term on the right-hand side of 
Eq. \ref{eq:exp-series} to the multipole $X_{i,LM}^{\mr{ln}}$ of the lognormal field is:

\begin{multline}
X_{i,LM}^{\mr{(2)}} = \\\sum_{l,m,l',m'}\frac{Z_{i,lm}Z_{i,l'm'}}{2}
\int Y_{lm}(\hat{\bm{\theta}})Y_{l'm'}(\hat{\bm{\theta}})Y_{LM}^*(\hat{\bm{\theta}})\mr{d}^2\hat{\theta}=\\
\sum_{l,m,l',m'}\frac{Z_{i,lm}Z_{i,l'm'}}{2}\sqrt{\frac{(2l+1)(2l'+1)(2L+1)}{4\pi}}\times\\
(-1)^M
\left(\begin{array}{ccc}
l & l' & L \\
0 & 0  & 0 \\
\end{array}\right)
\left(\begin{array}{ccc}
l & l' & L \\
m & m' & -M \\
\end{array}\right),
\label{eq:g-multipole-convol}
\end{multline}
where $\left(\begin{array}{ccc} l_1 & l_2 & l_3 \\m_1 & m_2 & m_3 \\ \end{array}\right)$ are Wigner 3-$j$ 
symbols. These can be non-zero if $m_1+m_2+m_3=0$ and $|l_1-l_2|\leq l_3\leq l_1+l_2$, which shows that 
$X_{i,LM}^{\mr{(2)}}$ can get contributions from $Z_{i,lm}$ with $l>L$ (see Fig. \ref{fig:3j-symbol}). 
The practical consequence of such non-locality is that if one wants to accurately simulate lognormal 
fields up to a bandlimit $L_{\mr{max}}$, it is necessary to generate Gaussian multipoles up to $l_{\mr{max}}>L_{\mr{max}}$. 
This fact is a general characteristic of lognormal fields and does not depend on the chosen transform 
(e.g. an analogous relation exists for Fourier transforms). Fig. \ref{fig:ln-power-loss} shows an 
example of this effect for the density contrast angular power spectra at redshift $z=0.2$ for a 
$\mr{\Lambda CDM}$ model; the larger the non-Gaussianity, the larger the effect. 

Lastly, the necessity of truncating the series in Eq. \ref{eq:legendre-trafo-xi} at a finite $\ell$ introduces 
a hard bandlimit that translates into oscillations in $\xi^{ij}(\theta)$. These oscillations can be 
minimised by increasing the series range to higher $\ell$ and/or by introducing a high-$\ell$ 
suppression in $C^{ij}(\ell)$. {\sc Flask} has the option of applying exponential suppressions and 
those that result from convolving the field with Gaussian and/or Healpix pixel window functions.

\subsubsection{Generating correlated multipoles}
\label{sec:alm-gen}

The statistical isotropy of the simulations allows for each multipole $Z_{i,\ell m}$ with different $\ell$ and $m$ 
indices to be generated independently. However, multipoles with the same $\ell$ and $m$ but from different 
fields and/or redshift slices (hence different $i$ indices) might be correlated. To generate such correlated 
Gaussian multipoles, we construct a covariance matrix for each multipole $\ell$ using the values of 
$C_{\mr{g}}^{ij}(\ell)$ as elements of that matrix. We then apply a Cholesky decomposition [using the GNU 
Scientific Library (GSL) routine\footnote{\texttt{\footnotesize{http://www.gnu.org/software/gsl}}}] 
to the covariance matrix $\bm{\mr{C_{\mr{g}}}}(\ell)$:

\begin{equation}
C_{\mr{g}}^{ij}(\ell) = \sum_k T_{ik}(\ell)T_{jk}(\ell),
\label{eq:cholesky}
\end{equation}
where $\mat{T}(\ell)$ are lower triangular matrices which can be used to generate the correlated Gaussian multipoles 
from a set of standard (zero mean and unit variance) independent Gaussian variables $Z^0_{k,\ell m}$:

\begin{equation}
Z_{i,\ell m}=\sum_k T_{ik}(\ell)Z^0_{k, \ell m}.
\label{eq:alm-gen}
\end{equation}
The computation of the expected value $\langle Z_{i,\ell m}Z_{j,\ell' m'} \rangle$ shows that $Z_{i,\ell m}$ 
indeed follow $C_{\mr{g}}^{ij}(\ell)\delta_{\ell\ell'}\delta_{mm'}$ where $\delta_{ab}$ are Kronecker deltas.

Non-positive-definite matrices (those with non-positive eigenvalues) cannot be decomposed as in 
Eq. \ref{eq:cholesky}; in fact, these are invalid covariance matrices in the sense that no set of 
variables can possibly have such covariances. Even though one can start with a set of $C_{\mr{ln}}^{ij}(\ell)$s 
that, as expected, leads to a positive-definite matrix $\mat{C_{\mr{ln}}}(\ell)$ for each $\ell$, the matrix 
$\mat{C_{\mr{g}}}(\ell)$ obtained as described in Sec. \ref{sec:ln2g-Cl} might not be positive-definite for two reasons: 
numerical errors and the fundamental limitation described in Sec. \ref{sec:harmonic-space}. While numerical errors are 
small enough such that a fractional change of $\la 10^{-4}$ in the $C_{\mr{ln}}^{ij}(\ell)$s 
(a negligible change for most cosmological applications) is sufficient to solve the problem, the second 
reason might need significantly larger fractional changes; moreover, it cannot be overcome with more 
accurate computations or different simulation methods since it is intrinsic to lognormal variables.

For generic fields, {\sc flask} can fix this problem by distorting the covariance matrices $\mat{C_{\mr{g}}}(\ell)$ as little 
as possible so they become positive-definite. Unfortunately, there are different ways to quantify the distance between 
two matrices; therefore, two methods are provided: one is guaranteed to minimise the Frobenius 
norm (quadratic sum of the matrix elements) of the difference between the regularised matrix and the original one 
\citep{Higham88mn}; and the other aims at applying the smallest fractional change possible to the matrix. The first 
one is quite fast since it simply performs an eigendecomposition of the matrix 
[$\mat{C_{\mr{g}}}(\ell) = \mat{Q}(\ell)\mat{\Lambda}(\ell)\mat{Q}(\ell)^{-1}$, where $\mat{Q}(\ell)$ 
is a matrix whose columns are eigenvectors of 
$\mat{C_{\mr{g}}}(\ell)$ and $\mat{\Lambda}(\ell)$ is a diagonal matrix formed by $\mat{C_{\mr{g}}}(\ell)$ 
eigenvalues] and then sets the negative eigenvalues to zero (or close to zero for numerical reasons). 

The second method is an iterative one and therefore takes more time. For an $N\times N$ matrix, it tries to obtain 
positive eigenvalues by applying successive fractional changes to its elements in the $N\times N$-dimensional 
direction of greatest change for the negative eigenvalues. In other words, it computes the gradient of the sum of the 
negative eigenvalues as a function of all matrix elements and follows it until all eigenvalues are positive. 
Although there is no rigorous proof that this method reaches the minimum fractional change required to make the 
matrix positive-definite, this is what can be expected from following the negative eigenvalues gradient and it indeed 
performs better in this sense than alternative methods like simply adding small values to the matrix diagonal or like 
the first one presented (however it is possible that the larger fractional changes produced by the first 
method might only affect uninteresting $C^{ij}_{\mr{g}}(\ell)$s with very low power). 

\subsubsection{Map generation}
\label{sec:map-gen}

Once the multipoles $Z_{i,\ell m}$ for the zero mean Gaussian fields are generated, we build {\sc healpix} maps 
$Z_i(\hat{\bm{\theta}})$ from them using the {\sc alm2map} {\sc healpix} function. 

If the goal is to generate Gaussian fields, no extra step is needed. However, if one wants to generate lognormal 
fields $X_i(\hat{\bm{\theta}})$, we have to apply the following local transformation to the Gaussian maps 
$Z_i(\hat{\bm{\theta}})$:

\begin{equation}
X_i(\hat{\bm{\theta}}) = e^{\mu_i}e^{Z_i(\hat{\bm{\theta}})}-\lambda_i,
\label{eq:g2ln-maps}
\end{equation}

\begin{equation}
e^{\mu_i} = (\langle X_i\rangle+\lambda_i)e^{-\sigma_i^2/2},
\label{eq:practical-mu}
\end{equation}
where $\sigma_i^2$ is the variance of the Gaussian field $Z_i(\hat{\bm{\theta}})$, given by 
\begin{equation}
\sigma_i^2 = \sum_{\ell=\ell_{\mr{min}}}^{\ell_{\mr{max}}}\frac{2\ell+1}{4\pi}C_{\mr{g}}^{ii}(\ell).
\label{eq:var-from-Cl}
\end{equation}
Although $e^{\mu_i}$ can also be directly related to the lognormal field variance 
(Eq. \ref{eq:mu-from-moments}) -- which in turn is related to $C_{\mr{ln}}^{ii}(\ell)$ 
by an equation identical to Eq. \ref{eq:var-from-Cl} -- the fact that in practice we 
generate Gaussian multipoles in the strict range $\ell_{\mr{min}}\leq\ell\leq\ell_{\mr{max}}$ 
means that the lognormal field multipole range is not well defined (see Sec. \ref{sec:ln2g-Cl}), 
making such calculation more difficult.

\begin{figure}
  \includegraphics[width=1\columnwidth]{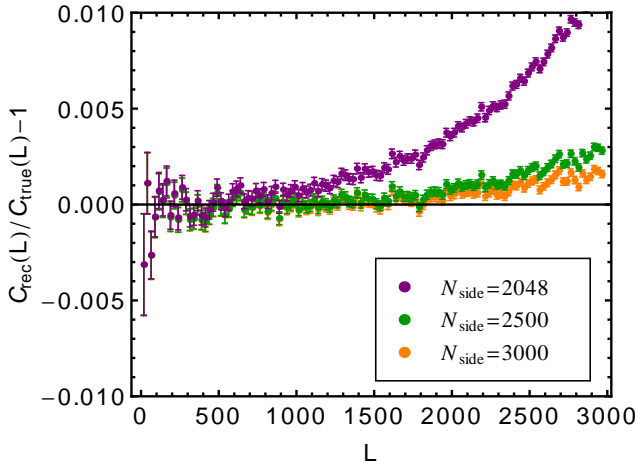}
  \caption{Fractional difference between the average of lognormal field power spectra measured 
    from 400 realizations of lognormal field maps and the original power spectrum, averaged inside 
    25-$\ell$ bins, for different map resolutions: $N_{\mr{side}}=2048$ (purple), $N_{\mr{side}}=2500$ 
    (green) and $N_{\mr{side}}=3000$ (orange). All maps were generated by exponentiating 
    Gaussian fields band-limited to $\ell_{\mr{max}}=7000$. The error bars represent the error on the mean.}
\label{fig:nside-effects}  
\end{figure}

The multipole mixing referred to in Sec. \ref{sec:ln2g-Cl} also introduces the issue that while 
the Gaussian field is band limited to $\ell_{\mr{max}}$, the exponentiation via Eq. \ref{eq:g2ln-maps} 
excites modes beyond $\ell_{\mr{max}}$ as Eq. \ref{eq:g-multipole-convol} exemplifies. Such an increase 
in the bandlimit is analogous to what happens in trigonometric identities such as 
$\cos^2(\omega\theta)=1/2 + \cos(2\omega\theta)/2$. This leads to the 
need of higher {\sc healpix} map resolutions than would be expected from the Gaussian field bandlimit 
to avoid aliasing effects. An example of this is shown in Fig. \ref{fig:nside-effects}, where we 
compare the original power spectrum with the ones reconstructed from full-sky lognormal maps with different resolutions, 
all generated from an associated Gaussian field with bandlimit $\ell_{\mr{max}}=7000$. While the resolutions as small 
as $N_{\mr{side}}=\ell_{\mr{max}}/4$ are enough for the Gaussian field, we need to go to 
$N_{\mr{side}}\simeq \ell_{\mr{max}}/3$ to get to one per cent precision on the lognormal field. Together with 
the need to simulate the Gaussian field up to higher multipoles than required, this makes lognormal 
field simulations more costly in terms of memory than Gaussian fields.   

Since homogeneity and isotropy of the fields on a spherical shell manifest themselves in the harmonic space 
as multipole independence $\langle a_{i,\ell m}a_{j,\ell'm'}\rangle = C^{ij}(\ell)\delta_{\ell\ell'}\delta_{mm'}$, 
one might ask if the multipole mixing that happens during exponentiation can break these symmetries 
by introducing correlations between different multipoles. Fortunately, this cannot happen since the transformation 
from a Gaussian to a lognormal field is local and homogeneous in real space over the spherical shell and 
as such cannot introduce a preferential location or direction. Nevertheless we verified the independence 
of the lognormal field multipoles by realising them 2000 times in the range $2<\ell<100$ and $0\leq m\leq\ell$ 
[since the field is real, $a_{\ell,-m}=(-1)^ma_{\ell,m}^*$] and measuring all 
their correlations [a total of $5148\times(5148+1)/2 = 13,253,526$ elements]. The off-diagonal terms 
($\langle a_{\ell m} a_{\ell' m'} \rangle$ with $\ell\neq\ell'$ and $m\neq m'$) only show random statistical 
fluctuations smaller than 5 per cent. This isotropy conservation means that the exponentiation process 
is a rotation (and possibly an isotropic dilation) in harmonic space.

\subsubsection{Density line-of-sight integration}

If a set of density maps were generated in contiguous redshift slices (of arbitrary thickness), 
{\sc flask} can use them to compute a convergence field for sources located at each slice boundary 
up to the last one at redshift $z_{\mr{last}}$ using the approximation described in Sec. \ref{sec:solutions}. 
For these convergence fields to be accurate, the redshift slices should be reasonably thin and 
cover the redshift range $0\la z<z_{\mr{last}}$. The power spectra expected to be followed by these 
convergence fields (e.g. the red curves in Fig. \ref{fig:los-int-recov-Cls}) can be computed by an 
auxiliary routine in {\sc flask}.  

\subsubsection{Shear computation}

Weak gravitational lensing caused by matter distributed along the LoS will distort the images 
of distant galaxies by bending the path travelled by the photons coming from them: if a photon 
comes from the true angular position $\hat{\bm{\beta}}$, it might be observed at a different position 
$\hat{\bm{\theta}}$. The distortion in the image can be described to first order by the partial derivatives of 
$\hat{\bm{\beta}}$ with respect to $\hat{\bm{\theta}}$ components, 
$\mr{\partial}\hat{\bm{\beta}}_i/\mr{\partial}\hat{\bm{\theta}}_j$, which is parametrized by the three parameters 
$\kappa$ (convergence), $\gamma_1$ and $\gamma_2$ \citep[shear components;][]{Bartelmann01mn}:

\begin{equation}
\frac{\mr{\partial} \hat{\bm{\beta}}_i}{\mr{\partial}\hat{\bm{\theta}}_j} = 
\left(\begin{array}{cc}
1-\kappa-\gamma_1 & -\gamma_2         \\
-\gamma_2         & 1-\kappa+\gamma_1 \\
\end{array}\right).
\label{eq:distortion-matrix}
\end{equation}
Both the shear distortion and the flux and apparent size magnification described by the convergence are 
caused by the same intervening matter, and the shear can be deduced from the convergence, whose angular power spectra should 
be provided by the user as input if one wants to generate shear fields (alternatively, the 
convergence can be computed from the density fields as presented in the previous section). The shear 
can be described as an expansion over spin-2 spherical harmonics ${}_{\pm2}Y_{\ell m}(\hat{\bm{\theta}})$
\citep[ as in {\sc healpix} convention]{Zaldarriaga97mn}: 

\begin{equation}
{}_{\pm}\gamma(\hat{\bm{\theta}}) = \gamma_1(\hat{\bm{\theta}})\pm i \gamma_2(\hat{\bm{\theta}}) = 
\sum_{\ell m}\gamma_{\ell m}{}_{\pm2}Y_{\ell m}(\hat{\bm{\theta}}).
\label{eq:spin-2-expansion}
\end{equation}   
According to \citet{Hu00mn}, on the full sky the harmonic multipoles $\gamma_{\ell m}$ are 
related to the convergence ones $\kappa_{\ell m}$ by:

\begin{equation}
\gamma_{\ell m} = -\sqrt{\frac{(\ell+2)(\ell-1)}{\ell(\ell+1)}}\kappa_{\ell m}.
\label{eq:gamma-kappa-relation}
\end{equation}   
Therefore the process of computing the shear $\gamma_1(\hat{\bm{\theta}})$ 
and $\gamma_2(\hat{\bm{\theta}})$ involves first computing $\gamma_{\ell m}$ from 
$\kappa_{\ell m}$. For Gaussian realisations, the latter are obtained directly 
as described in Sec. \ref{sec:alm-gen}, whereas for lognormal realisations they 
have to be extracted from the lognormal {\sc healpix} maps computed as described 
in Sec. \ref{sec:map-gen}. We then use the {\sc healpix} function {\sc alm2map\_spin} to transform 
the shear $E$ and $B$ mode multipoles, $E_{\ell m}$ and $B_{\ell m}$, into $\gamma_1(\hat{\bm{\theta}})$ 
and $\gamma_2(\hat{\bm{\theta}})$. According to the {\sc healpix} manual, $E_{\ell m}=-\gamma_{\ell m}$ 
and $B_{\ell m}=0$. 

In the lognormal case, we must obtain $\kappa_{\ell m}$ from the maps using the 
{\sc map2alm\_iter} function (with one iteration), and this process can introduce noise 
in the shears if the map resolution is too low compared to the shear bandwidth $\ell_{\mr{max}}$. 
To avoid that, we recommend $N_{\mr{side}}\sim\ell_{\mr{max}}$.

\subsubsection{Noise and selection effects}
\label{sec:sel-func}

Once the tracer density contrast fields $\delta_i(\hat{\bm{\theta}})$ are available, {\sc flask} 
can apply the survey selection function $\bar{n}_i(\hat{\bm{\theta}})$ (i.e. the expected observed density if the 
universe had no structure, provided by the user) to $\delta_i(\hat{\bm{\theta}})$ to get the 
expected observed density, used as the mean value of a Poisson distribution from which we will 
draw the actual observed tracer density $n_i(\hat{\bm{\theta}})$:

\begin{equation}
n_i(\hat{\bm{\theta}}) = \mr{Poisson}\{ \bar{n}_i(\hat{\bm{\theta}})[1+\delta_i(\hat{\bm{\theta}})] \}.
\label{eq:poisson-sampling}
\end{equation}   

The user has to provide a selection function for each one of the tracers (if there is more than one) and 
each selection function can be separable or not into angular and radial directions. For separable 
selection functions, the user must supply: a file describing the radial part as a table containing the 
redshifts and the expected number of observed tracers of that type per unit $\mr{arcmin^2}$ per unit redshift; 
and a {\sc healpix} map describing the fractions of this number that are observed at each angular coordinate. 
The final selection function is the product of these two. In the case of non-separable selection functions, 
a different {\sc healpix} map must be provided for each redshift slice, each one containing the expected 
number of observed tracers per unit $\mr{arcmin^2}$ per unit redshift.

\subsubsection{Catalogue building and output}

All quantities computed in the previous sections (from angular correlation functions $\xi^{ij}(\theta)$ 
in Sec. \ref{sec:ln2g-Cl} to {\sc healpix} maps of $n_i(\hat{\bm{\theta}})$ in Sec. \ref{sec:sel-func}) 
can be written to output files on request, along with other quantities like the $C_{\mr{ln}}^{ij}(\ell)$s 
obtained from the regularised $\mat{C_{\mr{g}}}(\ell)$ matrices described in Sec. \ref{sec:alm-gen} and 
the $C_{\mr{ln}}^{ij}(\ell)$s recovered from full-sky maps described in Sec. \ref{sec:map-gen}. The final 
{\sc flask} product and output is a catalogue of observed tracers that might contain the following 
columns, according to user request: angular position (using polar and azimuth angles or right ascension 
and declination, given in radians or degrees), redshift, tracer type, convergence, shear components, 
ellipticity components (see Eq. \ref{eq:ellipticity}), and a few bookkeeping numbers. 

Up to the catalogue creation step, all tracers inside a cell are associated with the cell's angular position 
(given by the {\sc healpix} map pixel centre position) and redshift (given by its redshift slice). 
During the catalogue creation process, each tracer in the cell gets a random angular position 
homogeneously sampled within the pixel boundaries and a random redshift sampled within its redshift 
slice according to an interpolation of the selection function such that, even if the simulated 
redshift slices are thick, the resulting radial distribution of tracers is smooth (no structure will 
be generated inside the slices, though). The ellipticities $\epsilon=\epsilon_1+i\epsilon_2$ are 
computed using the equation \citep{Bartelmann01mn}:

\begin{equation}
\epsilon = 
\begin{dcases}
\frac{\epsilon_{\mr{s}}+g}{1+g^{*}\epsilon_{\mr{s}}}, & |g|\leq 1; \\
\frac{1+g\epsilon_{\mr{s}}^{*}}{\epsilon_{\mr{s}}^{*}+g^{*}}, & |g|>1;\\
\end{dcases}
\label{eq:ellipticity}
\end{equation}
where $g\equiv {}_+\gamma/(1-\kappa)$ is the reduced shear and 
$\epsilon_{\mr{s}}=\epsilon_{\mr{s},1}+i\epsilon_{\mr{s},2}$ is the source intrinsic ellipticity, 
randomly drawn from a Gaussian distribution with variance set by the user. For introducing 
intrinsic alignment in the simulations, these have to be specified via the input power 
spectra.

\section{Summary and conclusions}
\label{sec:conclusions}

The lognormal modelling of large-scale structure (LSS) fields is an important tool for validating 
LSS data analysis, estimating covariance matrices and studying impact of noise, selection and 
systematic effects on the data. Current and future photometric LSS surveys require this modelling 
to be performed jointly for many observables, on the full sky and over a large redshift interval. 
In this paper we explained some of the obstacles faced by this task and described how the modelling can be 
performed accurately. We also presented a public code that can create such simulations for a wide 
range of field combinations.   

We showed in Sec. \ref{sec:toy-model} that lognormal variables cannot attain certain 
correlations or covariance matrices that would be, for instance, accessible to 
Gaussian variables. Although this is a known fact in the field of statistics, it remained 
unnoticed by the astrophysics community given it does not manifest itself when modelling 
density and convergence fields in an independent way, as was done so far. We then showed in Sec. 
\ref{sec:harmonic-space} that these limitations are propagated to the harmonic space 
in a similar but smoothed out way, and that as a consequence of 
such limitations the covariance matrix of the associated Gaussian variables becomes non-positive-definite. 
 
In Sec. \ref{sec:distortions} we presented a way of quantifying 
the amount of ``non-positive-definiteness'' by computing the fractional change required to 
make the covariance matrix positive-definite and showed that, when modelling both density and 
convergence as a multivariate lognormal field, the change required is much larger than the expected 
numerical errors, demonstrating that they are caused by intrinsic limitations in the model; 
therefore, better implementations of the same simulation process will not circumvent the problem.
We verified in Sec. \ref{sec:density-ln} that the multivariate lognormal model for both density 
and convergence is internally inconsistent as the density lognormality assumption leads to a 
non-lognormal distribution for the convergence. This is likely the reason why the lognormal model 
fails to result in valid covariance matrices for the associated Gaussian variables when modelling 
both fields together. 

We must therefore look for alternative methods if we want to create correlated random 
realisations of density and lensing. In this paper, 
two solutions were proposed: distorting the density and convergence auto and cross power spectra 
(Sec. \ref{sec:distortions}); or using non-lognormal convergence marginal distributions (Sec. \ref{sec:solutions}), 
for which we provided a fitting function in Sec. \ref{sec:pdf-fit}. Given that the lognormal model 
works well for the density and that the convergence can be obtained from 
the former by line-of-sight (LoS) integration, changing the shape of the convergence distribution to that 
of a sum of correlated lognormals allows both fields to be jointly modelled (see Fig. 
\ref{fig:los-int-recov-Cls}). Another (less attractive) possibility is to use the multivariate 
Gaussian model for creating realisations of the joint density and convergence fields: since the sum of 
Gaussian variables is also Gaussian, this model does not include the internal inconsistencies 
seen above. This alternative, however, has been shown to lead to underestimations of the convergence 
measurements covariance matrix \citep{Hilbert11mn}. 

Other bold possibilities would be to: (a) try different marginal distributions for the convergence 
[e.g. \citet{Das06mn, Schuhmann15mn} or different approximations to the sum of lognormal variables] 
or even for the density fields; or (b) try different copulas 
\citep[note that the multivariate lognormal distribution corresponds to the Gaussian copula with lognormal marginals;][]{Nelsen06x}. 
Compared to the multivariate 
lognormal distribution, both have the disadvantage that specifying the fields' statistical properties -- 
e.g. the power spectra -- might not be as straightforward as in the lognormal case (it might not even 
be analytically possible). Moreover, a different copula would still lead to the same limits in the 
fields' correlations described in Sec. \ref{sec:toy-model} given that they are limited by the 
Fr\'{e}chet--Hoeffding bounds.   

When modelling the convergence as a lognormal field, its shift parameter $\lambda$ (an additional parameter that specifies 
the minimum value of the lognormal distribution $X_{\mr{min}}=-\lambda$) 
is not fixed by the convergence power spectra and has to be determined 
somehow. Given that the true convergence field is not lognormal, different methods of estimating 
the shift parameter return different values (see Table \ref{tab:shift-methods}) that confer to the 
model different characteristics when compared to the real distribution.  
Using Eqs. \ref{eq:moments2shift} and \ref{eq:kappa-skew}, we provided a way to specify 
it directly from theory, without relying on ray-tracing simulations. In comparison to the method 
by \citet{Taruya02mn}, our method is more complex but it is built to reproduce the convergence skewness 
while the method of \citet{Taruya02mn} reproduces the minimum value observed in a finite sample.   
Given the arbitrariness on how $\lambda$ is set when modelling the convergence distribution, 
there is no reason to expect it to match the minimum value of the convergence 
(i.e. the one obtained in an empty LoS, $\kappa_{\mr{empty}}$) unless the fit is specifically built to reproduce this value 
in detriment of other properties. This conclusion leads us to question the common interpretation 
that the difference between $-\lambda$ and $\kappa_{\mr{empty}}$ is an indication that there are 
no empty lines of sight in the Universe. 

Finally, we presented in Sec. \ref{sec:code-description} the public code 
{\sc flask}\footnote{\texttt{\footnotesize{http://www.astro.iag.usp.br/\char`~flask}}} which 
is able to simulate an arbitrary number of correlated lognormal and Gaussian fields including 
multiple tracer densities, convergence and CMB radiation once 
their statistics are specified by an input power spectra set which can be computed by {\sc class} 
or {\sc camb sources}, for instance. In case the lognormal limitations prevent the realisation of the fields, 
{\sc flask} can overcome these limitations by distorting the input power spectra or by computing the 
convergence through a density LoS integration. Effects such as redshift space distortions, evolution, 
galaxy intrinsic alignments and arbitrary biases can be introduced by inscribing their effects 
in the power spectra, while selection functions and noise can be applied by {\sc flask} itself.  
The code adopts a tomographic approach on the full curved sky, thus making it ideal for large-area 
photometric surveys like DES, \emph{Euclid}, J-PAS, LSST and \emph{WISE}. 

\section{Acknowledgements}

The authors would like to thank Dr. Sreekumar Balan and Prof. Laerte Sodr\'{e} Jr. 
for useful discussions. This work was made possible by the financial support of 
FAPESP Brazilian funding agency. BJ acknowledges support by an STFC Ernest Rutherford 
Fellowship, grant reference ST/J004421/1.

\appendix

\section{Sum of correlated lognormal variables}
\label{sec:sum-lognormals}

We are interested in the first three central moments of the distribution of 
$Y$ which is a sum of lognormal variables $X_i$ weighted by $a_i$:
\begin{equation}
Y = \sum_ia_iX_i.
\label{eq:def-lognormal-sum}
\end{equation} 
They can all be easily computed from the one-, two- and three-point functions presented 
in Sec. \ref{sec:lognormal-variables}:
\begin{equation}
\langle Y\rangle = \sum_ia_i\langle X_i\rangle,
\label{eq:lognormal-sum-mean}
\end{equation} 

\begin{equation}
\langle Y^2\rangle-\langle Y\rangle^2 = \sum_{ij}a_ia_j\xi_{\mr{ln}}^{ij},
\label{eq:lognormal-sum-var}
\end{equation} 

\begin{equation}
\langle (Y- \langle Y\rangle)^3\rangle = \sum_{ijk}a_ia_ja_k\zeta_{\mr{ln}}^{ijk},
\label{eq:lognormal-sum-3mom}
\end{equation} 

\section{{\sc flask} usage example}
\label{sec:flask-example}

{\sc Flask} is executed in the command line by calling it together with a configuration file:

\noindent
{\tt flask sim-01.config}

\noindent
The configuration file specifies all settings using keywords followed by a colon. 
For instance, {\tt RNDSEED: 1243} specifies the random number generator seed and 
{\tt MAP\_OUT: data/kappa-map-01.dat} specifies an output file for a table of field 
values at all angular positions. {\sc Flask} comes with an example configuration file 
that describes all keywords; these are also explained in detail in {\sc flask}'s documentation. 

All settings can be overridden by providing new values in the command line, e.g.:

\noindent
{\tt flask example.config RNDSEED: 334 MAP\_OUT: ./map-002.dat}

\begin{figure}
  \includegraphics[width=1\columnwidth]{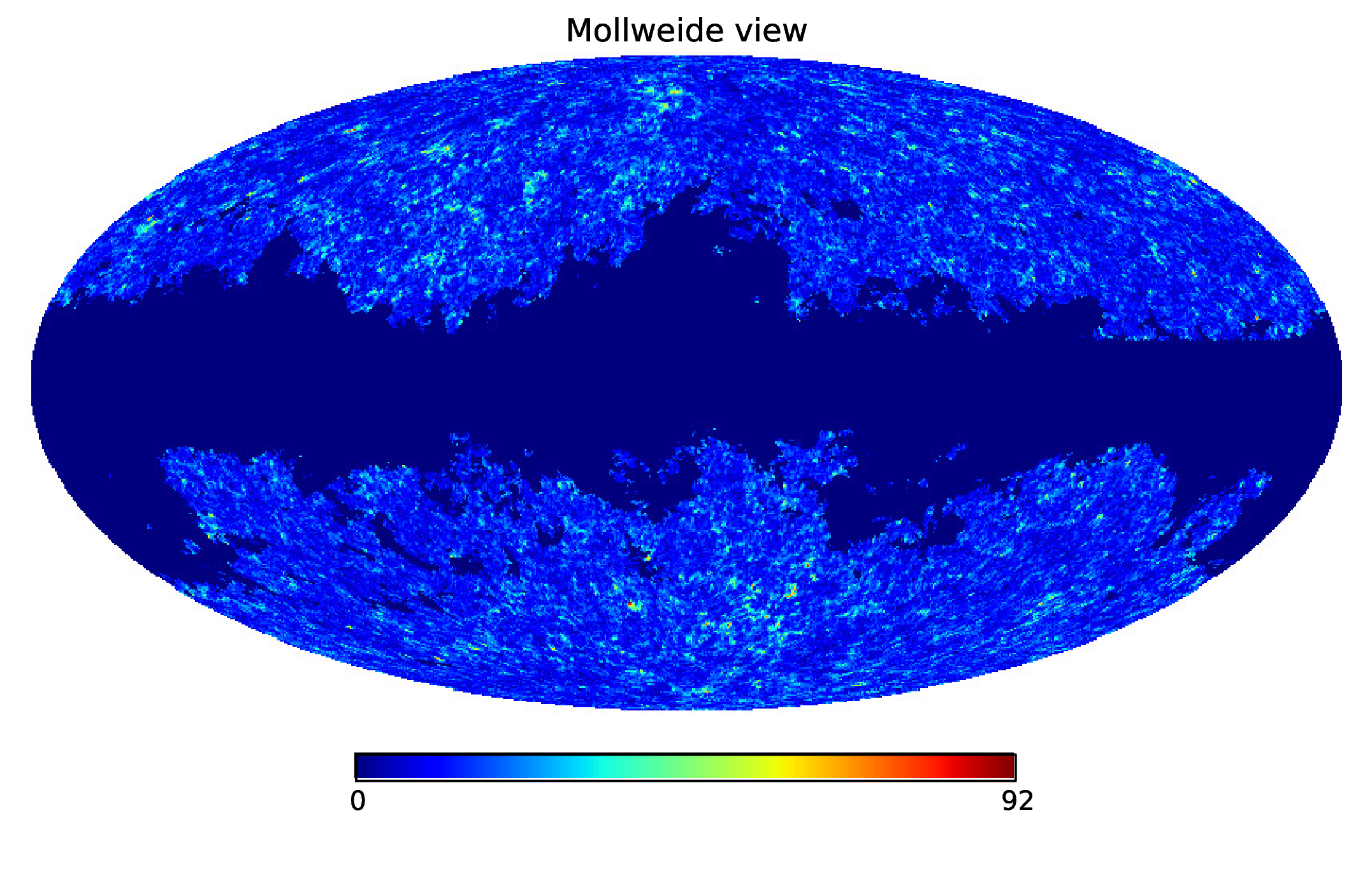}
  \caption{A Mollweide projection of a galaxy counts {\sc healpix} map produced by {\sc flask}, simulating the \emph{WISE} survey.}
\label{fig:wise-map}  
\end{figure}

Among many possible outputs, the code can produce {\sc healpix} maps of the generated fields. 
Fig. \ref{fig:wise-map} shows a {\sc flask} simulation of the \emph{WISE} survey 
(under a $\mr{\Lambda CDM}$ model), consisting of galaxy counts following 
the survey selection function, including the Milky Way angular mask.

\bibliographystyle{mn2e}
\bibliography{main}

\end{document}